\documentclass[12pt]{article}
\pdfoutput=1
\usepackage[utf8]{inputenc}
\usepackage{epsf}
\usepackage{colortbl}
\usepackage{amsmath}
\usepackage{amsfonts}
\usepackage{amssymb}
\usepackage{graphicx}
\usepackage{color}
\usepackage{psfrag}
\usepackage{cite}
\usepackage{hyperref}
\usepackage{tikz}
\usepackage{slashed}

\usepackage{ifpdf}

\definecolor{lightgray}{gray}{0.95}
\definecolor{lightblue}{RGB}{235,250,255}

\newcommand\greybox[1]{%
  \vskip\baselineskip%
  \par\noindent\colorbox{lightgray}{%
    \begin{minipage}{0.9\textwidth}#1\end{minipage}%
  }%
  \vskip\baselineskip%
}

\newcommand\bluebox[1]{%
  \vskip\baselineskip%
  \par\noindent\colorbox{lightblue}{%
    \begin{minipage}{0.9\textwidth}#1\end{minipage}%
  }%
  \vskip\baselineskip%
}

\newcommand{\bmat}{\left(\begin{array}}
\newcommand{\emat}{\end{array}\right)}

\def\yzero{\smash{\hbox{$y\kern-4pt\raise1pt\hbox{${}^\circ$}$}}}

\def\beq{\begin{equation}}
\def\eeq{\end{equation}}
\def\beqa{\begin{eqnarray}}
\def\eeqa{\end{eqnarray}}

\def\-{\hphantom{-}}

\def\s2{\frac{1}{\sqrt2}}

\def\beq{\begin{equation}}
\def\eeq{\end{equation}}
\def\beqa{\begin{eqnarray}}
\def\eeqa{\end{eqnarray}}

\def\IF{\relax{\rm I\kern-.18em F}}
\def\II{\relax{\rm I\kern-.18em I}}

\def\Dsl{\,\raise.15ex\hbox{/}\mkern-13.5mu D} 

\newcommand{\eq}[1]{(\ref{#1})}
\newcommand{\ket}[1]{\vert #1 \rangle}
\newcommand{\bra}[1]{\langle #1 \vert}

\catcode`\@=11   
\newdimen\@rotdimen
\newbox\@rotbox  

\def\@vspec#1{\special{ps:#1}}
\def\@rotstart#1{\@vspec{gsave currentpoint currentpoint translate
   #1 neg exch neg exch translate}}
\def\@rotfinish{\@vspec{currentpoint grestore moveto}}
\def\@rotr#1{\@rotdimen=\ht#1\advance\@rotdimen by\dp#1%
   \hbox to\@rotdimen{\hskip\ht#1\vbox to\wd#1{\@rotstart{90 rotate}%
   \box#1\vss}\hss}\@rotfinish}
\def\@rotl#1{\@rotdimen=\ht#1\advance\@rotdimen by\dp#1%
   \hbox to\@rotdimen{\vbox to\wd#1{\vskip\wd#1\@rotstart{270 rotate}%
   \box#1\vss}\hss}\@rotfinish}%
\def\@rotu#1{\@rotdimen=\ht#1\advance\@rotdimen by\dp#1%
   \hbox to\wd#1{\hskip\wd#1\vbox to\@rotdimen{\vskip\@rotdimen
   \@rotstart{-1 dup scale}\box#1\vss}\hss}\@rotfinish}%
\def\@rotf#1{\hbox to\wd#1{\hskip\wd#1\@rotstart{-1 1 scale}%
   \box#1\hss}\@rotfinish}%
\def\rotate{\@ifnextchar[{\@rotate}{\@rotate[l]}}
\def\@rotate[#1]#2{\setbox\@rotbox=\hbox{#2}\@nameuse{@rot#1}\@rotbox}

\catcode`\@=12

\topmargin
-1.5cm
\textwidth
15.5cm
\textheight
23.5cm
\oddsidemargin
0.7cm
\evensidemargin
1.2cm
\begin{document}

\makeatletter
\@addtoreset{equation}{section}
\makeatother
\renewcommand{\theequation}{\thesection.\arabic{equation}}
\hypersetup{pageanchor=false}
\pagestyle{empty}
\rightline{ }

\vspace{3cm}

\begin{center}
\LARGE{\bf A Holographic Derivation of the Weak Gravity Conjecture \\[12mm] }
\large{M. Montero\\[4mm]}
\footnotesize{
Instituut voor Theoretische Fysica, KU Leuven,\\
Celestijnenlaan 200D, B-3001 Leuven, Belgium
}

\vspace*{5mm}

\small{\bf Abstract} \\
\end{center}
\begin{center}
\begin{minipage}[h]{17.0cm}
The Weak Gravity Conjecture (WGC) demands the existence of superextremal particles in any consistent quantum theory of gravity. The standard lore is that these particles are introduced to ensure that extremal black holes are either unstable or marginally stable, but it is not clear what is wrong if this doesn't happen. This note shows that, for a generic Einstein quantum theory of gravity in AdS,  exactly stability of extremal black branes is in tension with rigorously proven quantum information theorems about entanglement entropy. Avoiding the contradiction leads to a nonperturbative version of the WGC, which reduces to the usual statement at weak coupling. The argument is general, and it does not rely on either supersymmetry or a particular UV completion, assuming only the validity of Einsteinian gravity, effective field theory, and holography.  The pathology is related to the development of an infinite throat in the near-horizon region of the extremal solutions, which suggests a connection to the ER=EPR proposal. 
\end{minipage}
\end{center}
\newpage
\hypersetup{pageanchor=true}
\setcounter{page}{1}
\pagestyle{plain}
\renewcommand{\thefootnote}{\arabic{footnote}}
\setcounter{footnote}{0}

\tableofcontents 

\vspace*{1cm}

\section{Introduction}
Perhaps one of the few universal things that we think we know about quantum gravity is that we cannot couple it to anything we want. You cannot just write down your favorite effective field theory, couple it to Einsteinian gravity, and hope that has a sensible UV completion.  Most such bottom up constructions are doomed to fail; this state of affairs has been colourfully called the Swampland of effective field theories \cite{Vafa:2005ui} (see also the review \cite{Brennan:2017rbf}), where there are only a few dry spots where UV-sensible theories live. If correct, the Swampland is awesome news, because  the sensible effective field theories outside of it are constrained. These constraints should be universal, low-energy predictions of quantum gravity, that one can check without building a collider the size of the Universe. This was recently illustrated in \cite{Reece:2018zvv}, where it was shown that the Swampland as we currently understand it predicts an exactly massless photon. 

So far, there is a healthy number of proposed Swampland criteria. Among these, one may count the Swampland Distance Conjecture \cite{Ooguri:2006in}, the Chern-Simons Pandemic \cite{Montero:2017yja}, or the recent conjecture on the scalar potential\cite{Obied:2018sgi,Agrawal:2018own}, which leads to the dramatic (and debated \cite{Andriot:2018wzk,Achucarro:2018vey,Garg:2018reu,Kehagias:2018uem,Dias:2018ngv,Denef:2018etk,Colgain:2018wgk,Roupec:2018mbn,Andriot:2018ept,Matsui:2018bsy,Ben-Dayan:2018mhe,Heisenberg:2018yae,Conlon:2018eyr,Kinney:2018nny,Dasgupta:2018rtp,Cicoli:2018kdo,Kachru:2018aqn,Murayama:2018lie,Akrami:2018ylq,Heisenberg:2018rdu,Choi18,Brahma18,Ashoorioon:2018sqb}) conclusion that there are no de Sitter vacua in string theory (this has been suspected long before \cite{Sethi:2017phn,Obied:2018sgi}; see \cite{Danielsson:2018ztv} for a thorough review of the stringy arguments against de Sitter, and \cite{Ooguri:2018wrx} for arguments supporting the conjecture). Most of these remain conjectures primarily supported by a large variety of stringy examples. There are few cases where more solid arguments can be given:  Absence of continuous spin representations can be rigorously proven in string theory \cite{Font:2013hia}, and there are also heuristic arguments supporting it \cite{Wigner:1963wwt}.  The absence of global symmetries in quantum gravity \cite{Abbott:1989jw,Coleman:1989zu,Kallosh:1995hi,Banks:2010zn} is very well supported by both heuristics \cite{Susskind:1995da}, worldsheet \cite{Banks:1988yz} and holographic \cite{Beem:2014zpa,Montero:2017mdq,Harlow:2018jwu,Harlow:2018tng} arguments.

One of the most far-reaching Swampland constraints is the Weak Gravity Conjecture (WGC) \cite{ArkaniHamed:2006dz}. In its original form, it says that in any effective field theory with gauge fields consistently coupled to Einsteinian gravity, there should be a state whose charge-to-mass ratio is equal to or larger than that of an extremal black hole. This has the effect of rendering the black hole unstable via Schwinger pair production of charged particles. This is only the mildest form of the conjecture. There are several more refined and constraining versions, such as a version applying to axions and other $p$-form fields, the Convex Hull Condition \cite{Cheung:2014vva}, Lattice and Tower WGC \cite{Heidenreich:2016aqi,Andriolo:2018lvp}, scalar versions of the conjecture \cite{Klaewer:2016kiy,Palti:2017elp,Lust:2017wrl,Landete:2018kqf} (leading to a connection to the Distance Conjecture and emergence \cite{Heidenreich:2017sim,Grimm:2018ohb,Heidenreich:2018kpg,Grimm:2018cpv}), and a variant leading to the instability of nonsupersymmetric AdS space \cite{Ooguri:2016pdq,Freivogel:2016qwc} (and to proposed constraints on the Standard Model \cite{Ibanez:2017oqr,Ibanez:2017kvh,Gonzalo:2018tpb,Gonzalo:2018dxi}). Variants of the WGC have also been argued to constrain axion models of inflation or relaxation \cite{delaFuente:2014aca,Rudelius:2014wla,Rudelius:2015xta,Montero:2015ofa,Brown:2015iha,Bachlechner:2015qja,Hebecker:2015rya,Brown:2015lia,Junghans:2015hba,Palti:2015xra,Heidenreich:2015nta,Kooner:2015rza,Heidenreich:2015wga,Ibanez:2015fcv,Montero:2016tif,Heidenreich:2016aqi,Hebecker:2016dsw,Saraswat:2016eaz,Herraez:2016dxn,Ooguri:2016pdq,Cottrell:2016bty,Hebecker:2017wsu}. The vanishing photon mass prediction mentioned above involves the WGC as well.

All of this makes the WGC one of the most interesting Swampland constraints, but why should it be true? On top of the fact that it seems to be satisfied in string compactifications, sometimes in very nontrivial ways \cite{Lee:2018urn,Lee:2018spm,Bonnefoy:2018mqb}, there is a heuristic argument that extremal black holes should be unstable \cite{ArkaniHamed:2006dz,Banks:2006mm}. Otherwise, one could take the $g\rightarrow0$ limit at finite $M_P$, leading to a theory with a global symmetry in quantum gravity, which is illegal \cite{Harlow:2018jwu,Harlow:2018tng}. This argument is unsatisfactory for a number of reasons. First, perhaps this dangerous limit cannot be taken. The Swampland Distance Conjecture suggests that points where $g\rightarrow0$ are generically near the boundary of the moduli space, where an infinite tower of fields become light and the effective field theory analysis breaks down. In fact, trying to follow the heuristics carefully in the AdS case leads precisely to this conclusion, and not to the WGC \cite{Montero:2017mdq}. On top of this, at any finite value of $g$, there is only a finite number of exactly stable black holes, so it is not clear how is this a problem. Theories with extended supersymmetry can have a large number of exactly stable extremal black holes with different charges \cite{Sen:2007qy}, and this seems to be fine. 

The goal of this paper is to provide a more convincing derivation of the WGC from more fundamental principles, a ``physics proof'' of sorts. This is not the first paper claiming to do so. Of course, the validity of any proof depends on the assumptions and axioms one is willing to accept. Reference \cite{Hod:2017uqc} argued that the WGC follows from a previous conjecture, the ``universal relaxation bound'', so it is only a proof if one is willing to count the latter as an axiom, or if one is able to prove it in turn from other axioms; see also \cite{Urbano:2018kax}. References \cite{Cottrell:2016bty,Shiu:2017toy} explored loop corrections to entanglement entropy in extremal black holes, finding no blatant pathologies, while reference \cite{Fisher:2017dbc} reached the opposite conclusion. This disagreement persists, with \cite{Andriolo:2018lvp} claiming that \cite{Fisher:2017dbc} uses certain loop correction formulae outside of their regime of validity. Reference \cite{Cheung:2018cwt} makes a  convincing case that higher-derivative corrections work in such a way so as to have small black holes satisfy the WGC, provided some assumptions about the UV -- more concretely, that it is described by an effective field theory such that the IR higher-derivative interactions arise as a result of classically integrating out UV modes --. These assumptions are not  general -- for instance, one of the original examples of the WGC in \cite{ArkaniHamed:2006dz} involves a non-supersymmetric $U(1)$ gauge field in 10d heterotic string theory, which does not have a field theory UV completion --, but when they hold, the argument is solid. Reference \cite{Hamada:2018dde} connects the WGC to causality and unitarity IR properties of the effective field theory. 

Just as the examples above, the argument in this paper is only as good as its assumptions. In this case, we consider a consider a Einsteinian quantum theory of gravity, but only in Anti-de Sitter (AdS) space, plus a number of technical assumptions discussed below. Of course, the point is that we now have the tools of holography at our disposal -- we can attack the problem from two sides, the bulk and the dual field theory description. A similar approach was followed in \cite{Harlow:2015lma,Montero:2016tif} and in \cite{Cottrell:2017ayj}, where we tried to relate the WGC to Lloyd's bound for the $\mathcal{C}=\mathcal{A}$ proposal for holographic complexity. See \cite{Nakayama:2015hga} for a thorough analysis of WGC bounds in AdS space. We will provide a derivation of the formula in Section IV C of that reference. 

I want to emphasize that ``holography'' as used here is a very mild assumption -- any consistent theory of gravity in AdS should have boundary correlators that obey unitarity and crossing symmetry, and is therefore a CFT \cite{Rychkov:2016iqz}.  Leveraging general theorems in the CFT side into bulk statements can be a powerful strategy, since unlike in other approaches one makes no explicit assumptions about the UV in the bulk --whether it is String Theory or something else, whether there is supersymetry or not, the arguments work all the same--. The argument given in this note applies to any bulk theory, as long as it reduces to Einsteinian gravity plus effective field theory at low energies. 

I will now briefly outline the argument. An Einstein -- $U(1)$ -- AdS theory contains stable extremal planar black brane solutions \cite{Chamblin:1999tk}. These have a finite area horizon and zero temperature. They are holographically dual to the (highly degenerate) ground state of the dual CFT, deformed by a chemical potential. The maximal extension of these black branes is dual to the corresponding thermofield double state. The bulk description allows us to compute a variety of properties of this thermofield double, in particular its entanglement entropy (via the Ryu-Takayanagi formula \cite{Ryu:2006bv}). This satisfies a volume law (this means that it grows as the volume enclosed by the entangling surface). 

At the same time, the extremal black brane has a near-horizon $AdS_2\times \mathbb{R}^{d-1}$ geometry, which controls the behavior of correlators in the IR. In particular, there is a notion of near-horizon scaling dimension, and a corresponding dilatation operator. These scaling dimensions are related to masses and charges of the states in the theory and, under mild assumptions, they lead to correlators that decay exponentially in space. Pure quantum states with exponential decay of correlations are widely expected to obey an area law \cite{Eisert:2008ur}.  In fact, this has been proven in a particular case \cite{Brandao:2014ppa,2013NatPh...9..721B}, which covers (a particular compactification of) the extremal black brane, which should therefore obey an area law, leading to a contradiction with the holographic behavior. It follows that the bulk model we are using must be inconsistent. To avoid the contradiction, one can modify the theory in such a way that the scaling dimensions no longer lead to exponential decay of correlations.  The simplest way to do this is to introduce a charged state whose charge is bigger than or equal to its mass - in other words, a WGC particle. This both covers the supersymmetric and nonsupersymmetric cases. In the supersymmetric case, the BPS particle makes the dilatation operator gapless. In the nonsupersymmetric case, the WGC particle introduces a near-horizon instability, which is seen as complex near-horizon scaling dimensions. In both cases, exponential decay of correlations no longer holds.

The pathological behavior in the entanglement entropy is directly related to the presence of a horizon even at zero temperature, which entails a large entropy density. In bottom-up holography this has long been regarded as probably unphysical, since there is no obvious symmetry protecting this degeneracy (see e.g. \cite{Natsuume:2014sfa}\footnote{Therefore we have two different communities (bottom-up holography and string model building) that get to the same conclusion (the WGC) by different routes. This connection seems to have been first noticed in \cite{Denef:2009tp}, as well as \cite{Barbon:2008sr} which discusses entanglement entropy, of more direct relevance to us.}). On the other hand, it was difficult to pinpoint a precise inconsistency; perhaps the finite entropy density of the extremal black brane was not unphysical, but just weird. The argument presented here aims to fill in this gap, pointing out that there is indeed a pathology if WGC particles are absent. In spirit, it follows the same line as \cite{Cottrell:2016bty,Shiu:2017toy}, which also tried to find a pathology in the entanglement entropy of extremal charged black holes in WGC-violating theories; the crucial difference is that holography allows us to leverage rigorous quantum information theorems into bulk statements. 

While it reduces to the usual WGC in perturbative setups, the requirement that the spectrum of the dilatation generator should not be holomorphic and gapped makes sense more generally. One might then regard this statement as a nonperturbative generalization of the WGC, a concrete property that consistent quantum theories of gravity must satisfy and that reduces to the ordinary WGC in particular cases.

The statement derived here is also connected to the ER=EPR proposal \cite{Maldacena:2013xja}. The zero-temperature thermofield double states dual to the black brane are highly entangled, yet the wormhole at zero temperature grows a logarithmic throat and effectively disconnects. This is is what causes the pathological volume law for the entanglement entropy, and at the same time, is in contradiction with ER=EPR, which demands a wormhole. The WGC introduces particles which either are so light that can traverse the wormhole in a finite time, even at zero temperature, or are tachyonic, causing an instability and the wormhole to fragment into a myriad of Planckian sized wormholes. 

\subsection{Outline of the paper}
This paper has a single goal - to provide a holographic derivation of the Weak Gravity Conjecture. The different Sections have been structured accordingly:\begin{itemize}
\item Section \ref{sec:pre} introduces the necessary preliminaries, which can be safely skipped by the familiar reader:
\begin{itemize}
\item Subsection \ref{sec:bh} focuses on (charged) black holes and how to relate them to CFT quantities. 
\item Subsection \ref{sec:nhtfd} discusses the thermofield double formalism and near-horizon symmetries of extremal black holes.
\item Subsection \ref{sec:EE} is a lightning review on entanglement entropy (EE) and holographic entanglement entropy computations. It also discusses the holographic EE computation for the black holes of interest.
\end{itemize}
\item Section \ref{sec:core} is the core of the paper: 
\begin{itemize} \item In Subsection \ref{sec:bhp}, a contradiction is presented between the holographic EE calculation and a quantum information theorem. 
\item Subsection \ref{sec:wgc} then presents the main result coming out of the previous Subsection, discussing how it reduces to the standard WGC in both supersymmetric and nonsupersymmetric cases. 
\end{itemize}
\item Finally, Section \ref{conclus} sums up results and conclusions.
\end{itemize}

\section{Preliminaries}\label{sec:pre}
The main result of the paper relies on black hole thermodynamics and holographic entanglement entropy calculations. While the result is general, it is useful to focus on a particular ``benchmark'' model, which in our case will be Einstein-Maxwell. We will also allow for a finite number of charged elementary particles, as long as their charge-to-mass ratio is low enough so that they do not satisfy the WGC. This Section reviews the necessary background.  The material is standard and can be safely skipped by a familiar reader. 

\subsection{AdS black holes and their thermodynamics}\label{sec:bh}
Our arena will be $(d+1)$-dimensional  Anti de Sitter space in the Poincar\'{e} patch. All the metrics we will consider will therefore asymptote to 
\begin{align}ds^2_{\text{AdS}}= \left(\frac{\ell^2}{r^2}\right) dr^2+\frac{r^2}{\ell^2}\left(-dt^2+\sum_{i=1}^{d-1}dx_i^2\right),\label{adsmetric}\end{align}
where $\ell$ is the AdS radius. The Poincar\'{e} horizon sits at $r=0$. As is well known \cite{Witten:1998qj,Chamblin:1999tk}, the conformal boundary of \eq{adsmetric}, which sits at $r\rightarrow\infty$, is $\mathbb{R}^d$. As usual the gravity theory in the Poincar\'{e} patch is dual to a conformal field theory on $\mathbb{R}^d$ endowed with the flat Minkowski metric.  

To proceed further, we need to specify the bulk theory. The minimum requirements for a WGC discussion are Einsteinian gravity (the WGC does not seem to hold in higher-spin theories \cite{Nakayama:2015hga}  such as Vasiliev's theory \cite{Giombi:2016ejx}) and a weakly coupled $U(1)$ gauge field. As usual, the $U(1)$ gauge field is dual to a conserved charge $Q$ in the boundary. By Einsteinian gravity we actually mean that there is a weakly coupled bulk effective field theory with a finite number of light fields of spin $<2$ \cite{Heemskerk:2009pn,ElShowk:2011ag}, as well as a decoupling limit for the gravitational sector.  We will refer to this as the large $N$ limit, as usual; however, we emphasize that we assume nothing but the existence of the limit. The parameter $N$ can therefore be discrete or continuous; we will take $G_N$ to scale as
\begin{align}c_T\propto G^{-1}_N\propto N^2,\end{align}
since this is the correct scaling in the more familiar $d=4$ case; but $N$ is just a bookkeeping device and our results will be general. Similarly, the $U(1)$ gauge coupling will scale as \cite{Nakayama:2015hga}
\begin{equation}g\sim 1/N\end{equation}
This scaling ensures that both gravity and gauge interactions become weak in the large $N$ limit but the product $g/\sqrt{G_N}$, which shows up in the WGC bound, remains finite. All the results mentioned in this paper are semiclassical, to leading order in $N$.

The strategy in this paper is to obtain the WGC by contradiction: we will start with a theory in which extremal black holes are stable, and derive a paradox. While the arguments we present are general, it is useful to keep a simple, concrete model in mind as a ``benchmark'' to illustrate the arguments. We will take the Einstein-Maxwell system with action
\begin{align}\int d^{d+1}x\left[\frac{1}{2\kappa_{d+1}^2}\left(R+\frac{d(d-1)}{2\ell^2}\right)-\frac{1}{4g^2} F_{\mu\nu}F^{\mu\nu}\right],\quad \kappa_{d+1}^2=8\pi G.\label{emac}\end{align}

The next step is to look for black hole solutions of \eq{emac} which asymptote to \eq{adsmetric}. These are worked out in \cite{Chamblin:1999tk}. The metric is 
\begin{align}ds^2=-U(r)dt^2+ \frac{dr^2}{U(r)}+\frac{r^2}{\ell^2}\left(\sum_{i=1}^{d-1}dx_i^2\right),\quad U(r)\equiv\frac{r^2}{\ell^2}-\frac{m}{r^{d-2}}+\frac{q^2}{r^{2d-4}}.\label{bbmetric}\end{align}
There is a horizon at the larger solution of $U(r)=0$. We will call this solution $r_+$. To avoid naked singularities, one must impose the (sub)extremality condition,
\begin{equation} \frac{d}{d-2} r_+^{2d-2}\geq q^2\ell^2.\end{equation}
There is also an electric field \cite{Chamblin:1999tk,Montero:2017mdq},
\begin{align}A=\mu-\frac{1}{c}\frac{\sqrt{2}g}{\kappa_{d+1}}\frac{q}{r^{d-2}} dt, \quad c\equiv\sqrt{\frac{2(d-2)}{d-1}}.\label{bbgfield}\end{align}

These are ``black branes'', since the topology of the horizon is $\mathbb{R}^d$. They can be obtained from standard AdS Reissner-Nordstrom black hole solutions via a scaling limit \cite{Chamblin:1999tk}. Since the solutions extend to infinity in the transverse directions, one should talk of mass and charge densities. These correspond to energy and charge density in the corresponding CFT state, given by \cite{Chamblin:1999tk,Montero:2017mdq}
\begin{align}M=\frac{(d-1)}{16\pi G} m,\quad Q=\frac{\sqrt{(d-2)(d-1)}}{\sqrt{8\pi G} g} q.\label{mcd}\end{align}
Since the solutions have a horizon, there is also a nonvanishing entropy density, proportional to the horizon area:
\begin{align}S=\frac{A}{4G}=\frac{2\pi r_+^{d-1}}{\kappa_{d+1}}.\label{entrobb}\end{align}
Finally, the solution \eq{bbmetric} has an Euclidean counterpart, obtained by Wick-rotating the time coordinate and introducing a periodicity in the time coordinate of $2\pi \beta$, where
\begin{align}\beta=\frac{4\pi\ell^2r_+}{dr_+^2-(d-2)(c\,\Phi \ell)^2},\quad \Phi \equiv \frac{\kappa_{d+1}}{\sqrt{2}g}\mu.\label{betaf}\end{align}
is the Hawking temperature of the horizon. As is well known \cite{Witten:1998qj,Witten:1998zw,Chamblin:1999tk}, the Euclidean solution encodes the thermodynamic properties of the Lorentzian one, and of the CFT in general. More specifically, in the semiclassical limit one can compute the partition function of the CFT at an inverse temperature $\beta$ and chemical potential $\mu$
\begin{align}Z_{CFT}=e^{-\beta G}=\sum_{\text{Saddles}} e^{-S},\label{sum-over-saddles}\end{align}
where $S$ is the Euclidean action of the solution and the sum ranges over solutions of the Euclidean equations of motion with the correct asymptotics for the bulk electrostatic potential,
\begin{align}A(r\rightarrow\infty)\rightarrow\mu+\ldots\end{align}
There is only one solution of the form \eq{bbmetric} for given $\beta$ and $\mu$. Then, \eq{sum-over-saddles} collapses to a single term, and we can identify the free energy of the CFT with $I/\beta$, where $I$ is the on-shell action of the black hole solution per unit transverse volume. In short, the bulk dual of the CFT at finite temperature and chemical potential is the corresponding black hole \eq{bbmetric}. We will be particularly interested in the small $T$ region. At $T=0$, the entropy contribution disappears, and the only contribution comes from the ground states of the modified CFT Hamiltonian $H_\mu= H-\mu Q$. In the bulk, this corresponds to an extremal black brane. 

Thus, we arrive at the picture that the extremal black brane is describing the ground state of the CFT modified by a chemical potential. Since this has a nonzero entropy density \eq{entrobb}, we expect this ground state to be highly degenerate, with the black brane describing a CFT mixed state which is a statistical superposition of all ground states with the same weight. The dimension of the ground state space grows exponentially with the volume of the system. 

In supersymmetric black branes, we understand this huge degeneracy. The theory has BPS operators, whose dimension is protected, and black holes/branes are built out of these. In the black hole case, one just has to count how many possible combinations are there for a particular value of the total charge \cite{Strominger:1996sh}. This is the standard success story of entropy counting in supersymmetric black holes. In short, the degeneracy is explained by supersymmetry. But \eq{entrobb} applies for non-supersymmetric black holes as well. We have a huge ground state degeneracy, and no explanation for it \cite{Barbon:2008sr}. It has been suggested that this is an artifact of the large $N$ limit \cite{Hartnoll:2016apf}, but there is nothing going wrong in an obvious way, and the limit does work in the supersymmetric case. In a way, the argument in Section \ref{sec:core} will explain what is wrong with this situation, in terms of entanglement entropy, while also being compatible with the supersymmetric result.

Since the equivalence between the black brane and the CFT ground state is one of the main ingredients in the argument, it is important to understand exactly when it holds. Equation \eq{sum-over-saddles} is a standard saddle point approximation in the large $N$ limit. Specifically, since the action \eq{emac} scales like $N^2$, the path integral localizes to the sum over saddles \eq{sum-over-saddles}. For each term in the sum, one then uses the usual  formula \cite{Denef:2009kn}, 
\begin{equation}\int e^{-N^2S}\propto \frac{1}{\sqrt{\text{det}\, S''}}e^{-N^2S_0},\end{equation}
where $S_0$ is the value of the action at the saddle point $\sqrt{\text{det}\, S''}$ is the determinant of the quadratic fluctuations around the saddle. The saddle point calculation will be reliable as long as $\text{det}\, S''>0$. A negative determinant signals the presence of a tachyonic instability in the black hole background. If it is zero, validity of the saddle point approximation depends on higher-order corrections (in the finite-dimensional case, this is studied by catastrophe theory \cite{poston1996catastrophe}.).

The exact conditions for instability of a probe scalar field in the background \eq{bbmetric} have been studied extensively; the Einstein-Maxwell system together with a charged scalar provides the simplest model of a holographic superconductor. In this case, the effective mass of the charged scalar is shifted by the electric field of the black hole, and there will be an instability if and only if this effective mass is below the near-horizon $AdS_2$ BF bound \cite{Gubser:2008px}.  As worked out in \cite{Denef:2009tp,Nakayama:2015hga}, this constraint that one must satisfy to ensure the absence of instabilities is essentially (the opposite of) the WGC: The charge-to-mass ratio of any field on the black hole background is bounded from above. As long as this happens, the saddle point approximation is reliable, even at extremality, since the contribution coming from the perturbations never becomes larger than the classical one. 

 We will therefore assume $\text{det}\, S''>0$ for any $\mu,T$ we consider; equivalently, black holes are (perturbatively) stable. If they are not, one just has to look at the effectively massless/tachyonic direction around the Euclidean saddle; that mode is a WGC particle. 

Similarly, the requirement that the Euclidean black brane is nonperturbatively stable is equivalent to demanding that \eq{bbmetric} is the dominant saddle in \eq{sum-over-saddles}. There are certainly no other saddles within the consistent truncation \eq{emac}, and one also does not expect them outside of it within the EFT framework. In many cases, one can make a given supergravity solution be the dominant contribution to a certain path integral via changes in boundary conditions in AdS, which correspond to different ensembles in the CFT \cite{Chamblin:1999tk,Chamblin:1999hg}.

In short, we will assume that the black brane \eq{bbmetric} is the dominant saddle in the partition function. Extrapolating from the benchmark, this will constitute one of the main hypothesis of our argument - that we have some sort of extremal black brane solution with finite horizon area which is dual to the ground state of a suitable deformation of the CFT.

\subsection{Thermofield doubles and near-horizon geometry}\label{sec:nhtfd}

\begin{figure}[!htb]
\begin{center}
\includegraphics[height=5.5cm]{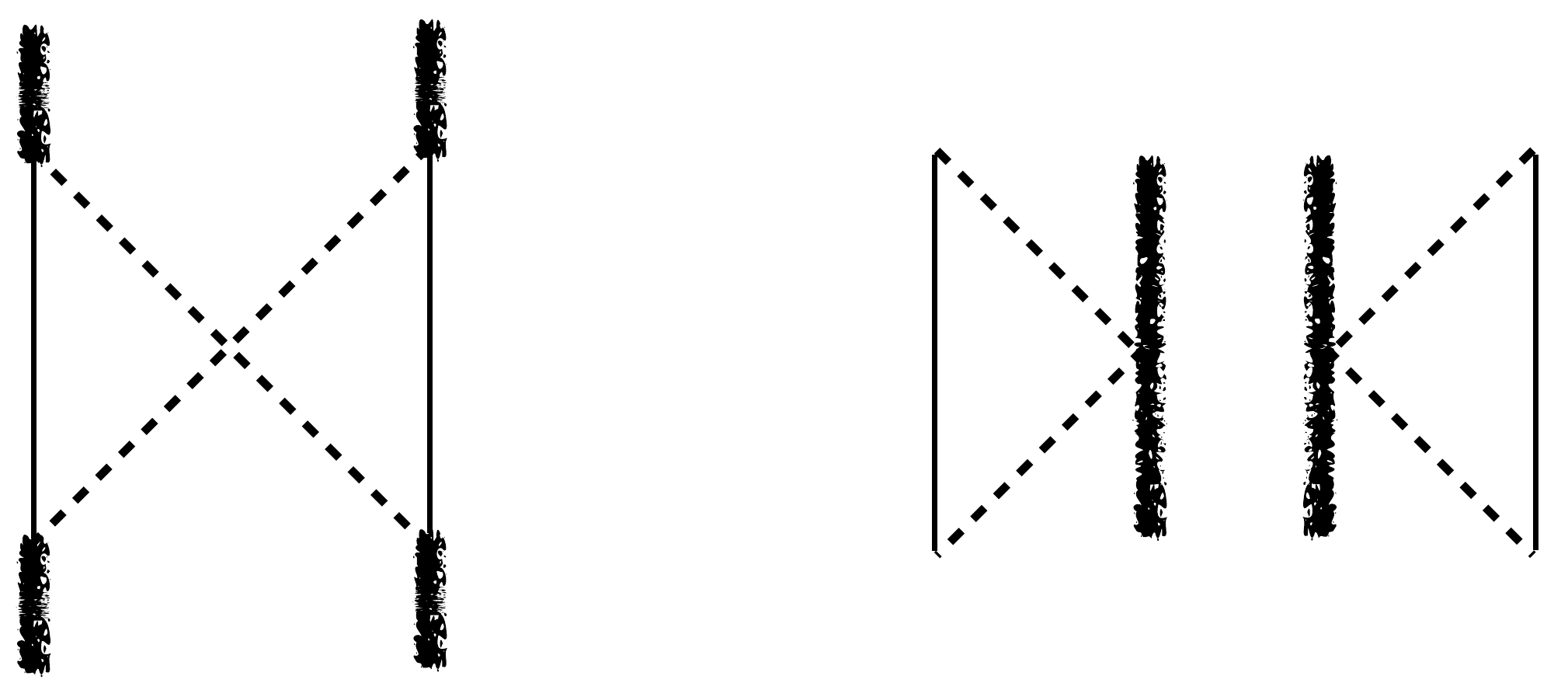}
\end{center}
\caption{On the left, conformal diagram of the maximal extension of the sub-extremal black brane. At each point we have ommited the transverse $\mathbb{R}^d$ factor. The solid lines represent the two conformal boundaries where each of the CFT copies in the thermofield double state \eq{tfd} live. Dashed lines represent black brane (outer and inner) horizons. Beyond the inner horizon lie timelike singularities, represented by fuzzy lines. When the black brane becomes extremal, the outer and inner horizons coincide; the left and right pieces disconnect, and the resulting conformal diagram is depicted on the right.}
\label{f2}
\end{figure}

The discussion in the previous Subsection involves the black brane only up to the horizon. However, the Lorentzian solution \eq{bbmetric} admits a continuation beyond, to a two-sided black brane. The conformal diagram is depicted in Figure \ref{f2}. As is well known, this now has two asymptotically AdS regions, so it is natural to guess that the holographic dual lives in two copies of the CFT. In fact, the holographic dual is supposed to be a thermofield double \cite{Maldacena:2001kr},
\begin{equation}\ket{TFD}=\frac{1}{\sqrt{Z(\beta,\mu)}}\sum_n e^{-\frac{\beta}{2}(E_n-\mu Q_n)}\ket{n}_L\ket{\bar{n}}_R\label{tfd}\end{equation}
an entangled state between the two CFT's. Here, $\ket{\bar{n}}$ means the CPT conjugate of $\ket{n}$. This state has the property that when tracing out one of the CFT's one is left with the thermal state for one CFT - which, as discussed in Section \ref{sec:pre}, is dual to a single-sided black hole. The two pictures agree, with the black hole interior somehow containing information about correlations of the two CFT's. 

\begin{figure}[!htb]
\begin{center}
\includegraphics[width=3.5cm]{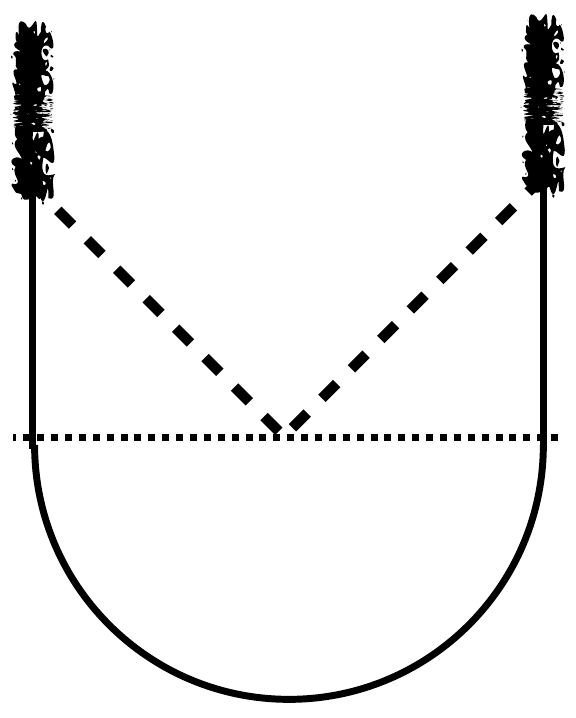}
\end{center}
\caption{Standard illustration of the Euclidean path integral preparing the charged thermofield double state \eq{tfd}. The Euclidean part of the solution is below the dotted line; its boundary is an interval of length $\beta/2$. The fuzzy lines represent the timelike singularities of the subextremal Reissner-Nordstron black brane. We have omitted the transverse $\mathbb{R}^{d-1}$ factor.}
\label{fb3}
\end{figure}

The thermofield double state can be prepared by a simple Euclidean path integral \cite{Maldacena:2001kr}: Just consider the vacuum CFT on an interval of length $\beta/2$, as illustrated in Figure \ref{fb3}. The corresponding bulk saddle is half of the Euclidean saddle contributing to the thermal partition function. 

We will be interested in the zero-temperature limit of \eq{tfd}, where the black brane becomes extremal. In this limit, the only states contributing to \eq{tfd} have minimal $E_n-\mu Q$, and as a result \eq{tfd} becomes an energy eigenstate in the dual CFT.  In this limit, \eq{tfd} involves only the groundstate subspace of the modified Hamiltonian $H_\mu=H-\mu Q$,
\begin{equation}\ket{TFD(T=0)}\propto\sum_{\text{Groundstates}}\ket{a}_L\ket{\bar{a}}_R\label{tfdt0}.\end{equation}

As we approach \eq{tfdt0} by lowering the temperature, an extra symmetry emerges deep in the bulk. Near the region where the two horizons intersect in Fig. \ref{f2}, the metric simplifies at low enough temperatures. To lowest order in $\beta^{-1}$, and introducing the coordinate $z\equiv r-r_+$, where $r_+$ is the extremal black brane radius, the function $U(r)$ in \eq{bbmetric} becomes
\begin{equation}U(r)\approx \frac{z^2}{\ell_{AdS_2}^2}+ 2\pi T\left(2z+(5-3d)\frac{z^2}{r_+}\right)+\mathcal{O}(T^2),\quad \frac{1}{\ell_{AdS_2}^{2}}\equiv\frac{(d-1)d}{\ell^2}. \label{nhads2}\end{equation}
The leading term in \eq{nhads2} is the dominant contribution for 
\begin{equation} 4\pi \ell_{AdS_2}^{2}T\ll z\ll r_+.\end{equation}
In this range, the metric \eq{bbmetric} is to a good approximation
\begin{equation} ds^2\approx -\frac{\ell_{AdS_2}^2}{z^2}dt^2+\frac{dz^2}{\ell_{AdS_2}^2}+r_+^2\left(\sum_{i=1}^{d-1} dx_i^2\right).\label{ads2metric}\end{equation}
This is the metric in the Poincar\'{e} patch of a product space $AdS_2\times\mathbb{R}^{d-1}$, where the $AdS_2$ factor involves the $(z,t)$ coordinates, and the transverse $\mathbb{R}^{d-1}$ factor with coordinates $x_i$ has a flat metric. 

Due to the near-horizon $AdS_2\times \mathbb{R}^{d-1}$ region, one could expect the ground state dual to the extremal black brane to develop an emergent $SL(2,\mathbb{R})$ symmetry (see e.g. \cite{Sen:2008yk,BrittoPacumio:1999ax}). Morally, there should be three conformal generators, obeying $SL(2,\mathbb{R})$ commutation relations, 
\begin{equation}i[D,H]=K,\quad i[D,K]=-H,\quad i[K,H]=D\label{so21algebra}\end{equation}
all of which should annihilate the vacuum. In the one-sided description, where the black brane represents a mixed ground state, this means that $[G,\rho]=0$ for each $G$ in \eq{so21algebra}. In the thermofield double picture, which describes both sides of the wormhole, this becomes the usual statement $G\ket{TFD}=0$. In the $\beta\rightarrow\infty$ limit, the Euclidean path integral defining the thermofield double state becomes the real line; the symmetry \eq{so21algebra} is just the conformal group acting on this real line.

It is tricky to argue that these isometries of the black brane solution uplift to full symmetries of the theory, so perhaps the algebra \eq{so21algebra} is not exactly realized. Yet weakly interacting particles in AdS, which correspond to single-trace operators in the dual CFT \cite{ElShowk:2011ag}, are sensitive only to the geometry and electric field, and indeed correlators of local operators behave as they should in a theory with $SL(2,\mathbb{R})$ symmetry. In particular, two-point functions in momentum space are of the form \cite{Hartnoll:2016apf}
\begin{equation} \langle \mathcal{O}(\vec{p},t)\mathcal{O}'(-\vec{p},t')\rangle\sim \frac{1}{\vert t-t'\vert^{2\Delta}},\end{equation}
which is what one would expect from conformal invariance. This property is one of the essential ingredients in the argument presented in Section \ref{sec:core}. Notice that the scaling coordinate is time only; spatial coordinates do not scale. It is an example of a $z=\infty$ Lifshitz scaling point \cite{Hartnoll:2016apf}. The IR tail of any correlator in the dual CFT in the black brane background is controlled by this $SL(2,\mathbb{R})$ symmetry. The conformal diagram splits in two in the $\beta=\infty$ case, as illustrated in Fig \ref{f2}, so we end up with two sets of independent near-horizon $AdS_2$ regions.  Accordingly, two-sided correlators vanish in this limit.

We should emphasize that the usual issue with $AdS_2$, namely that backreaction is so strong that any finite-energy excitations break the conformal symmetry \cite{Maldacena:2016upp}, is not expected to be present in this case due to the transverse $\mathbb{R}^{d-1}$ factor \cite{Almheiri:2014cka}. States with well-defined energy (which would have strong backreaction in the $AdS_2$ factor) are non-normalizable plane waves in the transverse $\mathbb{R}^{d-1}$. But for normalizable wavepackets this effect is damped. Relatedly, the thermodynamic gap of a black hole \cite{Almheiri:2016fws} vanishes as the size of the black hole diverges. 

To sum up, the thermofield double \eq{tfdt0} is described by a two-sided extremal black brane in the bulk, which enjoys an emergent $SL(2,\mathbb{R})$ symmetry due to its near-horizon geometry, at least as far as weakly interacting particles are concerned.

\subsection{(Holographic) entanglement entropy}\label{sec:EE}

The main point of this paper is to turn entanglement entropy theorems into Swampland constraints, so a somewhat detailed introduction to entanglement entropy is in order. The reader is encouraged to read the excellent review \cite{Rangamani:2016dms}, parts of which we will now roughly summarize.

Suppose one is given a density matrix $\rho$ - that is, a hermitean, unit trace, positive linear operator on some Hilbert space $\mathcal{H}$. One of the simples quantities that one can associate to $\rho$ is its von Neumann entropy
\begin{equation}S(\rho)\equiv-\text{Tr}\left(\rho\,\log\, \rho\right).\label{vneum}\end{equation}
One thing this quantity measures is how pure the state $\rho$ is. For a pure state, $\rho=\ket{\Psi}\bra{\Psi}$, $S(\rho)=0$, while it achieves its maximal value for $\rho\propto\mathbf{I}$. In case $\rho$ is a thermal state, $\rho=e^{-\beta H}/Z(\beta)$, $S$ is precisely the thermodynamic entropy. 

It turns out that \eq{vneum} also provides a measure of quantum entanglement. Roughly speaking, the word ``entanglement'' describes the presence of correlations in a quantum bipartite system that do not have a classical counterpart; there is no (local) classical system that could ever reproduce them \cite{Horodecki:2009zz}. Given a pure state $\ket{\Psi}$ in a bipartite system described by a tensor product Hilbert space $\mathcal{H}_A\otimes\mathcal{H}_B$, we would like to know whether $\ket{\Psi}$ is entangled and if so, how much. It turns out that the Von Neumann entropy of either reduced state
\begin{align}\rho_A=\text{Tr}_B(\ket{\Psi}\bra{\Psi})\quad\text{or}\quad \rho_B=\text{Tr}_A(\ket{\Psi}\bra{\Psi})\end{align}
is a measure of entanglement\footnote{In a precise sense \cite{Horodecki:2009zz}, this means that the entanglement entropy remains invariant under any ``classical'' operation, and therefore is a measure of purely quantum correlations.}, called the entanglement entropy of $\ket{\Psi}$. As a simple check, when the state $\ket{\Psi}$ factorizes as $\ket{\Psi}_A\otimes\ket{\Psi}_B$, so that there are no bipartite correlations, the entanglement entropy of the reduced state $\rho_A=\ket{\Psi}_A\bra{\Psi}_A$ indeed vanishes.

Notice that the entanglement entropy is an actual entanglement measure only when applied to pure states. Nevertheless, its definition makes sense for mixed states as well, where it also captures some of the classical correlations in the system. 

In a quantum field theory defined on $\mathbb{R}^{d}$ (such that states live naturally in $\mathbb{R}^{d-1}$), there is a natural class of bipartitions, given by 
\begin{align} \mathcal{R}&\equiv\textbf{A region of $\mathbb{R}^{d-1}$ with boundary $\mathcal{S}$},\nonumber\\ \mathcal{R}^c&\equiv\textbf{The complement of $\mathcal{R}$ in $\mathbb{R}^{d-1}$.}\end{align}
We will therefore talk of the entanglement entropy of the state $\ket{\Psi}$ (or a mixed state $\rho$) associated to the surface $\mathcal{S}$, and denote it by $S_{\ket{\Psi}}(\mathcal{S})$ or $S_{\rho}(\mathcal{S})$. One is often interested on how the entanglement entropy behaves as a function of the size of $\mathcal{S}$; sometimes we will replace $\mathcal{S}$ by some characteristic scale $L$ and talk about the entanglement entropy $S_{\ket{\Psi}}(L)$ as a function of $L$. To get meaningful statements, we need to introduce a regularization, since the entanglement entropy in field theory turns out to be UV divergent. The intuitive picture is that a generic state of a quantum field theory is entangled, so when tracing out one of the subsystems we get contributions from the degrees of freedom right across the entangling surface $\mathcal{S}$. But in a continuum field theory we get infinitely many degrees of freedom per unit volume, which causes a divergence. A simple way out is to e.g. introduce a lattice regularization. As the lattice spacing $\epsilon\rightarrow0$ one expects the entanglement to be dominated by modes just across the entangling surface $\mathcal{S}$, leading to what is called an area law for entanglement entropy:
\begin{equation}S_{\ket{\Psi}}(\mathcal{S})\rightarrow\frac{\mathcal{A}(\mathcal{S})}{\epsilon^{d-1}}\quad\text{as $\epsilon\rightarrow0$}.\end{equation}
As we will see, this universal behavior is automatically reproduced by holographic entanglement entropy.

Because of its universal character, the above area law does not tell us much about the state $\ket{\Psi}$ itself. It will be more interesting to us to look at the opposite regime, where the size of $\mathcal{S}\rightarrow\infty$ while keeping the cutoff $\epsilon$ finite and constant. The behavior of the entanglement entropy in this limit is captured by universal terms, which are independent of the cutoff. Alternatively, one can look at particular combinations of entanglement entropies, such as mutual information, in which the cutoff dependence cancels out. We will see an example of this in the next Section. 

For the rest of the paper we will be interested in the IR limit. Here the entanglement entropy can also display simple behaviors. Most states in a quantum field theory will satisfy a volume law: the entanglement entropy grows as the volume of the region enclosed by $\mathcal{S}$. Because we have introduced a lattice regularization, the Hilbert space associated to the region enclosed by $\mathcal{S}$ is actually finite-dimensional, and the logarithm of its dimension scales with the volume. Because the maximum Von Neumann entropy in a state space of dimension $d$ is $\log d$, it follows that volume law is the fastest entanglement entropy can grow. 

While a typical state in a QFT will have a volume law \cite{2006CMaPh.265...95H}, special states can have a different behavior. The most prominent example is the ground state of the theory, which often obeys an area law (see \cite{Eisert:2008ur}). This is very interesting on its own, since an area law means that the ground state can be analyzed via the techniques of matrix product states \cite{Eisert:2008ur}. However, it is by no means a universal statement: There are mild violations in two-dimensional CFT's \cite{Calabrese:2009qy}, which have a universal logarithmic behavior, and systems with Fermi surfaces often get an area law with an extra logarithmic factor \cite{Eisert:2008ur}. Furthermore, there are some exotic one-dimensional systems with polynomial or even volume law entropies in the ground state \cite{2010JMP....51b2101I,2014arXiv1408.1657M}. We will elaborate on the origin of the area law in Section \ref{sec:core}.

\begin{figure}[!htb]
\begin{center}
\includegraphics[width=6.7cm]{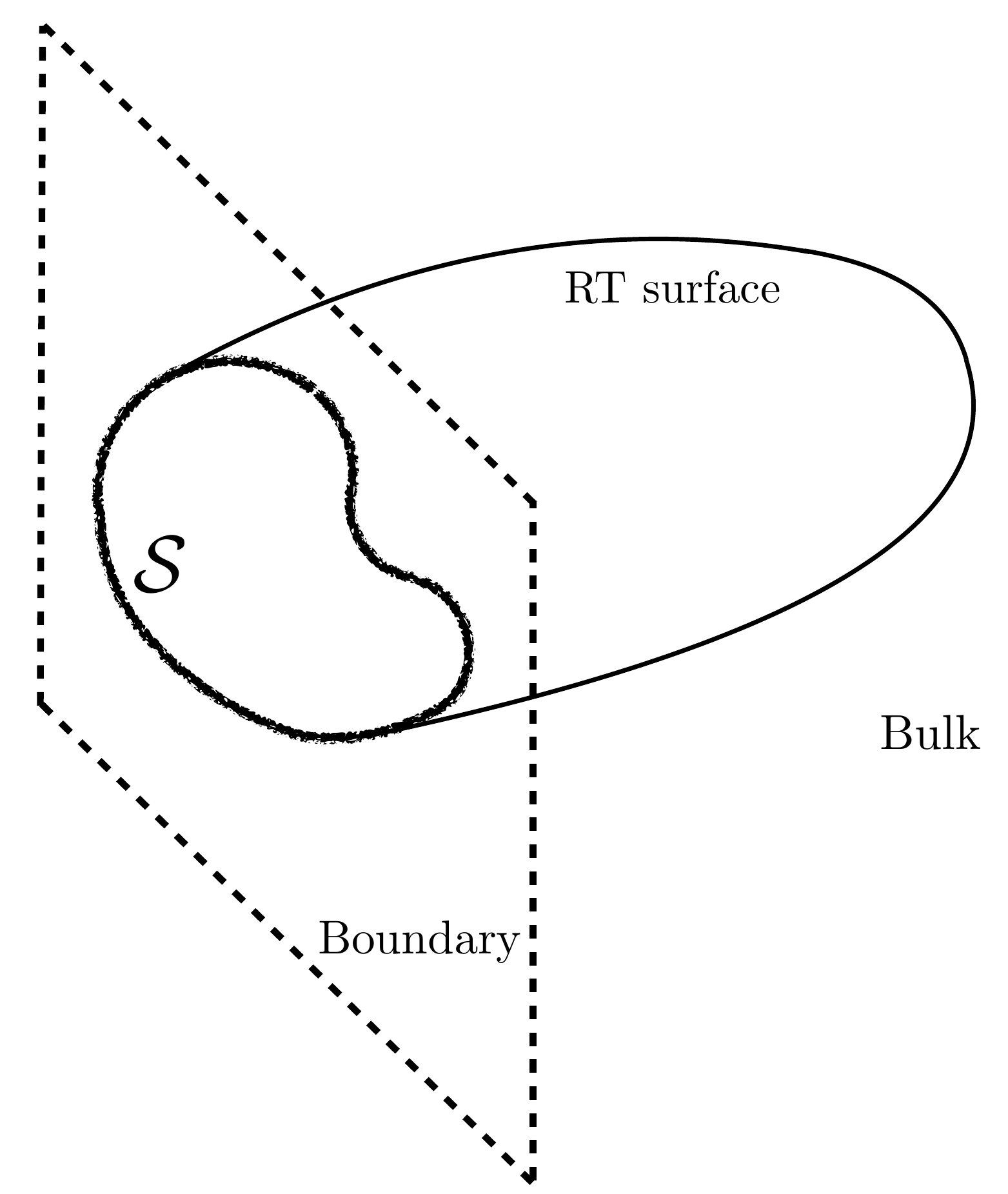}
\end{center}
\caption{Illustration of the Ryu-Takayanagi prescription for computing entanglement entropies. The dotted square represents a regularized version of the boundary, in which the entangling surface $\mathcal{S}$ sits. The RT surface is anchored in $\mathcal{S}$, extends into the bulk, and has minimal area.}
\label{f1}
\end{figure}

In a holographic CFT (by which again we mean a CFT with some sort of large $N$ limit in which gravity decouples and there is a low-energy effective field theory containing finitely many fields of spins $\leq 2$ \cite{Heemskerk:2009pn,ElShowk:2011ag}), Ryu and Takayanagi provided a simple recipe to compute entanglement entropies, illustrated in Figure \ref{f1}: Given an entangling surface $\mathcal{S}$, consider all (properly regularized) bulk surfaces that anchor in $\mathcal{S}$ and that are homologous to it. Among these surfaces, find the one of minimal area. The entanglement entropy of the CFT state dual to the bulk geometry under consideration is simply
\begin{equation}S_{\rho}(\mathcal{S})=\frac{2\pi\mathcal{A}}{\kappa_{d+1}}, \label{rtformula}\end{equation}
where $\mathcal{A}$ is the area of the extremal surface. 

The RT formula \eq{rtformula} works equally well for pure or mixed states, and was originally derived for the case when the state under consideration must be approximately dual to a semiclassical geometry\footnote{In \cite{Almheiri:2016blp} the formula was argued not to work for superpositions of many different semiclassical states, but \cite{Harlow:2016vwg} showed this can be corrected by appropriately taking into account the contribution of the bulk entanglement entropy.}. Also, it is a large $N$ result, providing only the leading piece of the boundary entanglement entropy. As usual in holographic setups, $1/N$ corrections are dual to quantum loop effects in the bulk and are organized in a power series expansion in $\kappa_{d+1}$ \cite{ElShowk:2011ag}.  With these caveats, however, the formula was proven in \cite{Lewkowycz:2013nqa}, and first quantum corrections were analyzed in \cite{Faulkner:2013ana} (see the review \cite{Rangamani:2016dms} for a summary of the proof as well). The idea is to compute the entanglement entropy via the replica trick \cite{Rangamani:2016dms} and then compute the action of the replicas via a bulk semiclassical gravity computation. 

Using \eq{rtformula}, one can compute the entanglement entropy for a wide variety of states of interest. For instance, in the $AdS_{d+1}$ vacuum, one obtains an area law, as advertised. The Schwarzschild black brane leads to a volume law, which fits right in with the fact that this describes a thermal state in the dual CFT. Quite generally, for large enough entangling surfaces $\mathcal{S}$ in spacetimes with horizon, the RT surface tends to hover right above the horizon (see \cite{Rangamani:2016dms} again for details), resulting in an entanglement entropy proportional to the Bekenstein-Hawking black hole entropy. Some sample calculations can be found in Appendix \ref{app:C}. For convenience, we will now quote the results for the black brane metric \eq{bbmetric} for the case when $\mathcal{S}$ is a $S^{d-2}$ sphere of radius $L$:
\begin{equation}S_{\rho}=\frac{2\pi \mathcal{A}}{\kappa_{d+1}} \approx  \frac{2\pi r_+^{d-1}}{\kappa_{d+1}} \frac{\omega_{d-2} L^{d-1}}{d-1},\label{lala}\end{equation}
This volume law is actually independent of the precise shape of $\mathcal{S}$, as long as it is large enough. 

This concludes the lightning review of holographic entanglement entropy. Before we go on, however, there is an important point we must address. We are interested in the behavior of entanglement entropy as the characteristic length scale of $\mathcal{S}$, call it $L$, diverges. On the other hand, \eq{rtformula} only allows us to compute entanglement entropies in the large $N$ limit. So we have an issue of commutativity of limits. Suppose the actual, exact entanglement entropy at finite $N$ and $L$ can be written as
\begin{equation}S(\rho)=N^2 f(L)+ g(L,N),\label{comulims}\end{equation}
where $g(L,N)$ grows slower than $N^2$ in the large $N$ limit at constant $L$. The physically relevant quantity to check whether we have area or volume law is obtained by taking large and fixed $N$ first, then sending $L\rightarrow\infty$, while what we are actually able to compute is the $N\rightarrow\infty$ result for any fixed large $L$ (the $f(L)$ piece in \eq{comulims}). 

It is very hard to imagine that the limits would not commute for a pure state. The derivation in \cite{Lewkowycz:2013nqa} shows that the RT prescription arises as a saddle point action in semiclassical gravity, which shows no pathology no matter how large $L$ is.  For the limits not to commute, there would have to be some quantum gravitational effect that becomes increasingly important as $L\rightarrow\infty$. This would be a breakdown of local effective field theory on long scales. And in fact, one could ask similar questions in all sort of holographic situations: Consider, for instance, the calculation of the partition function of a CFT, as a function of $\beta$. As $\beta\rightarrow\infty$, $Z\rightarrow 1$, since the only contribution comes from the ground state. This is exactly matched by the $AdS$ saddle in the bulk perspective. However, this saddle comes with its own set of $1/N$ corrections. How do we know that there is no correction that overcomes the semiclassical contribution as  $\beta\rightarrow\infty$?

To get around this kind of objections, we will make the additional assumption that effective field theory and hence the semiclassical result \eq{rtformula} actually gives the leading piece of the entanglement entropy for arbitrarily large $L$. A similar point already arose at the end of Subsection \ref{sec:bh}, and will be a recurring theme of the paper, so we will state it clearly:  We will assume validity of the low-energy effective field theory, which means that the leading behavior of all quantities of interest in this paper (entanglement entropies, heat capacity, etc) is correctly reproduced by the semiclassical calculation.

\section{The WGC meets quantum information theorems}\label{sec:core}
We are finally in a position to present the black hole information paradox that leads to a version of the WGC. This reformulation involves asymptotically AdS black branes instead of black holes, and covers both the supersymmetric and non-supersymmetric cases. 

\subsection{A black hole paradox}\label{sec:bhp}

Just as in the previous Section, we will consider a two-sided black brane which is dual to a zero-temperature thermofield state in two copies of the dual CFT. One can think of the electric black brane \eq{bbmetric} as a benchmark, but it will be clear that the story is more general - it follows through whenever one has a zero-temperature state with a horizon and an emerging $AdS_2$ geometry in the deep IR.

As reviewed in Section \ref{sec:pre}, under these circumstances we expect an emergent conformal IR symmetry. In particular, this symmetry controls correlators of Euclidean local single-trace operators $\langle \mathcal{O}(\vec{x},t)\mathcal{O}'(\vec{x'},t')\rangle$ in the regime \cite{Iqbal:2011in,Hartnoll:2016apf,Gralla:2018xoz}
\begin{equation}\vert \vec{x} -\vec{x'}\vert,\vert t-t'\vert \gg 1/\mu.\end{equation}
In particular, after Fourier-transforming in time and space, one obtains the behavior \cite{Faulkner:2009wj,Iqbal:2011in,Hartnoll:2016apf}
\begin{equation}\langle  \mathcal{O}(\vec{p},\omega)\mathcal{O}(-\vec{p},0)\rangle \propto \omega^{2(\Delta_{\mathcal{O}}(\vec{p})-1/2)}, \label{corrbeh} \end{equation}
where $\Delta_{\mathcal{O}'}(\vec{p})$ is the would-be near-horizon conformal dimension. This is the eigenvalue under $iD$ of the near-horizon conformal symmetry. 

One can Fourier-transform \eq{corrbeh} back into real space, to obtain a late-time behavior as $\vert t-t'\vert^{2\Delta(0)}$ at long but fixed spatial separation. As long as $\Delta(\vec{p})$ is holomorphic near the real axis, this leads to exponential decay of spatial correlations at finite time \cite{Iqbal:2011in}. The near-extremal geometry is the prototypical example of this behavior, which has been dubbed quasi-local quantum criticality \cite{Iqbal:2011in,Hartnoll:2016apf}, where correlations decay on spatial directions but are long lived in time. The fact that \eq{corrbeh} is not gapped for any $\vec{p}$ implies the existence of light modes at any value of the momentum, which provides a heuristic explanation of the large ground state degeneracy in \eq{entrobb} \cite{zaanen2015holographic}.

Notice that the exponential decay of correlations only follows from \eq{corrbeh} at sufficiently late times. However, as proven in Appendix \ref{app:lieb}, one can show that \eq{corrbeh} implies that correlations decay exponentially at large spacelike separation also at $t=0$ if the following two assumptions hold:
\begin{itemize}
\item $\Delta_{\mathcal{O}}(\vec{p})$ is holomorphic on a strip near the real line for every operator in the theory.
\item $\Delta_{\mathcal{O}}(\vec{p})>0$ for every operator.
\end{itemize}

\begin{figure}[!htb]
\begin{center}
\includegraphics[width=5cm]{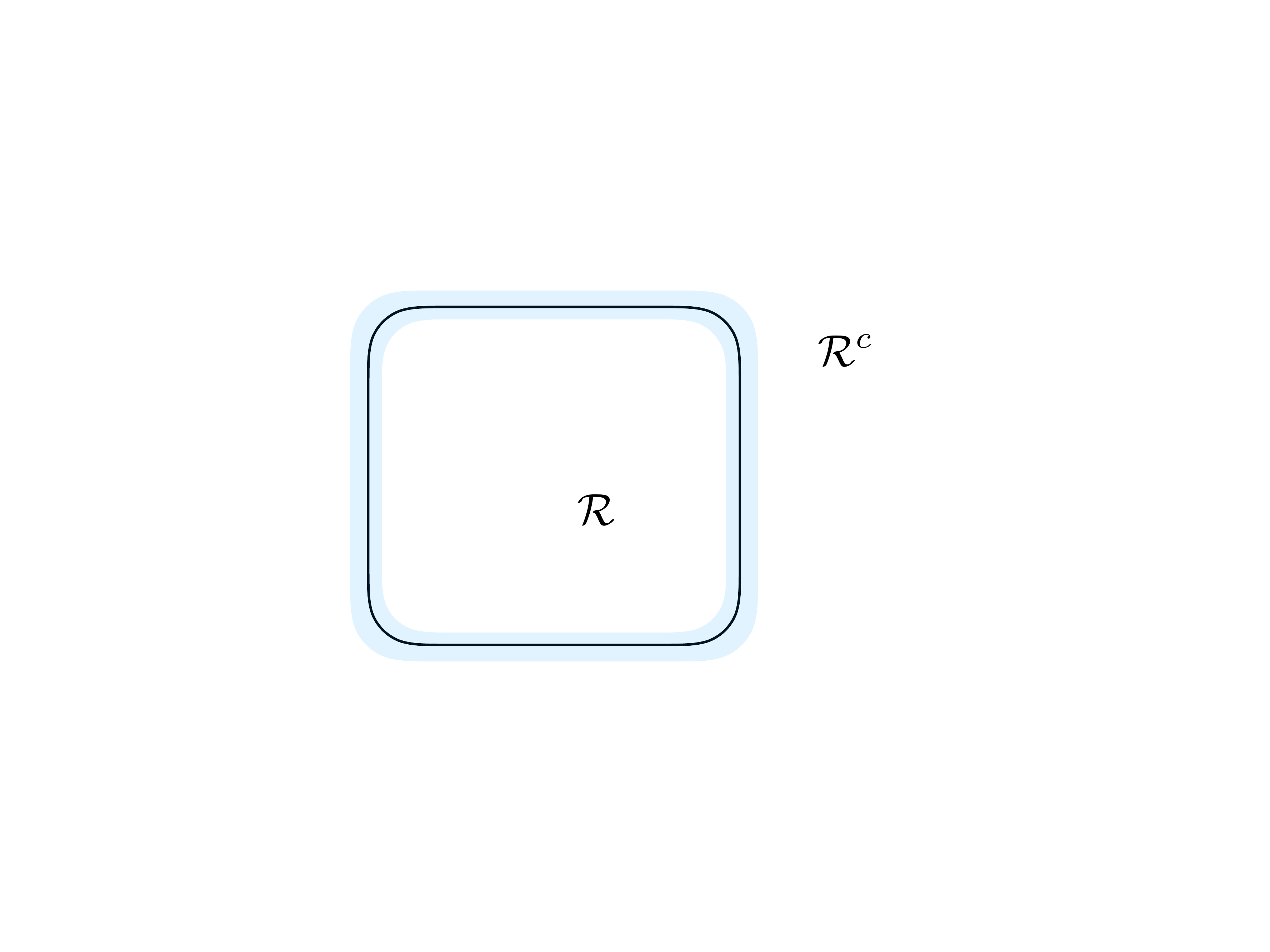}
\end{center}
\caption{In a system with exponential decay of correlations, one could intuitively expect the degrees of freedom in the bulk of region $\mathcal{R}$ to have almost no entanglement with degrees of freedom in $\mathcal{R}^c$. This would mean that only degrees of freedom in a neighbourhood of the boundary (shaded in blue in the picture) are significantly entangled, leading to an area law.}
\label{falw}
\end{figure}

From here on, we will take this two properties for granted, so we have a state with exponential decay of correlations for local operators. There is a broad expectation that such a behavior of correlators leads to an area law for the entanglement entropy \cite{Eisert:2008ur}. The heuristic explanation is illustrated in Figure \ref{falw}; the exponentially vanishing correlations intuitively mean that degrees of freedom in the bulk of region $\mathcal{R}$ are in fact unentangled with bulk degrees of freedom in region $\mathcal{R}^c$. In fact, this heuristic picture is what underlies the expectation that ground states of gapped Hamiltonian satisfy an area law (which by itself has been proven rigorously only in one dimension so far \cite{Hastings:2007iok}), since they always have exponential decay of correlations \cite{Hastings:2005pr}. 

This expectation is not correct in general: Quantum ``data-hiding'' states have tiny classical correlations, yet they can store large amounts of entanglement at the same time \cite{2001quant.ph..3098D}.  However, if it turns out to be correct in our setup, then we have a paradox: The extremal black brane shows exponential decay of correlations, which leads to an area law, which contradicts the volume law predicted by Ryu-Takayanagi. Subsection \ref{sec:wgc} discusses the way out of this contradiction, which essentially leads to the WGC. 

The remainder of this Subsection is devoted to making this handwavy ``paradox'' as sharp as we can\footnote{At first sight, a  similar paradox could be obtained from \cite{Brandao:2014uya}, which predicts a subvolume law for the entanglement entropy just from a linear heat capacity, which is the case for the extremal black brane. The argument would be even more direct, as there would be no need to check exponential decay of correlations. Unfortunately, the result in \cite{Brandao:2014uya} requires the unstated assumption that the ground state entropy is not extensive, which precludes a direct application. However, the argument could still work if one can prove that the thermofield double state is the ground state of a sufficiently local Hamiltonian; see \cite{Cottrell:2018ash} for recent progress in this direction.}. More concretely, the expectation above has been rigorously shown to be correct in the context of spin chain (1+1 dimensional lattice systems with a finite-dimensional local Hilbert space). We will first state the rigorous result, proven in \cite{Brandao:2014ppa,2013NatPh...9..721B}, and then discuss how to connect the black branes we are interested in to it. 

\begin{center}\greybox{\textbf{(Area law from exponential decay of correlations)} Consider a pure state $\ket{\Psi}$ in a one-dimensional qubit spin chain. Suppose there are $(l_0,\xi)$ such that for any operators $\mathcal{O}_{A}$ and $\mathcal{O}_{B}$ supported on intervals $A$ and $B$, separated $l>l_0$ sites, one has
\begin{equation}\langle \mathcal{O}_{A}\mathcal{O}_{B}\rangle_{\ket{\Psi}}\leq \vert\vert \mathcal{O}_{A}\vert \vert \vert\vert \mathcal{O}_{B}\vert \vert   2^{-l/\xi},\end{equation}
where $\vert\vert \cdot\vert \vert$ is an operator norm. Such a state is said to have exponential decay of correlations. Then, the entropy of a region of any size is bounded above by a constant,
\begin{equation}S_{\ket{\Psi}}(L)\leq c l_0 \exp(c'\xi \log\xi),\label{alawth}\end{equation}
where $c,c'$ are universal constants. In other words, the state satisfies an area law. 
}\end{center}

If we want to relate \eq{alawth} to the WGC, have to bridge a wide gap between a spin chain on one side and a holographic CFT on the other. The first obstacle that needs to be addressed is that \eq{alawth} applies to lattice systems, while we want to use it in a continuum  field theory. Bridging this gap requires the extra assumption that the CFT admits a lattice formulation, that is, there is a lattice model which reproduces CFT correlators in the deep IR. This lattice model provides a UV regulator of the CFT, which we also need to discuss entanglement entropies as reviewed in Section \ref{sec:EE}; the lattice spacing will be related to $1/\epsilon$.

Crucially, ``lattice model'' above does not just mean standard lattice spin systems, but also includes lattice gauge theories. This is an important extension, since for many well-known holographic CFT's we have lattice gauge theory descriptions, but no known spin lattice description\footnote{This does not mean that such a description does not exist, as 2d dualities illustrate \cite{Wegner:2014ixa,Mathur:2016cko}.}. Lattice gauge theories are not spin systems, but are obtained from them via a projection to the gauge-invariant subsector. Also, in a spin system, the degrees of freedom live in lattice sites, while a lattice gauge theory includes degrees of freedom (gauge fields) that live on lattice links \cite{Munster:2000ez}. These two quirks mean that the physical Hilbert space in a lattice gauge theory does not admit a local factorization (see e.g. \cite{Casini:2013rba}), which means that even the standard definition of entanglement entropy needs to be fixed. 

Here, we will follow the proposal set out in \cite{Ghosh:2015iwa}, which exploits the fact that while the physical gauge-invariant Hilbert space does not factorize, it is a subspace of the larger, unphysical Hilbert space, which does. One can then compute entanglement entropies in this enlarged space. The resulting entanglement entropy has a number of desirable properties, such as e.g. the replica trick. For our purposes, the main point is that the enlarged, non-gauge invariant system where the entropy is computed is an ordinary lattice system. The projection to the physical Hilbert space is imposed via a local constraint at each lattice site, which only involves nearest-neighbour degrees of freedom. As an example, in a $U(1)$ gauge theory with matter, gauge invariance is simply Gauss' law, $\vec{\nabla}\cdot\vec{E}=\rho$, point by point in spacetime. In the lattice gauge theory, this becomes the statement that physical operators are annihilated by
\begin{equation} \mathcal{G}_{\vec{x}}=\rho_{\vec{x}}+\sum_{\vec{\mu}\,\in\, \text{n.n}} \vec{u}_{\vec{\mu}}\cdot \left(\vec{E}_{\vec{x},\vec{x}+\vec{\mu}}-\vec{E}_{\vec{x},\vec{x}-\vec{\mu}}\right),\end{equation} 
for every $\vec{x}$ in the lattice. The sum $\vec{\mu}$ runs over all the nearest neighbours, and $\vec{u}_{\vec{\mu}}$ is the unit vector in the $\vec{\mu}$ direction. The lattice gauge theory also comes equipped by a gauge-invariant Hamiltonian (in the sense that $[H,\mathcal{G}_{\vec{x}}]=0$). The constrained lattice gauge theory system, satisfying the constraints $\mathcal{G}_{\vec{x}}\ket{\psi}=0$, can be described as the low-energy sector of the modified Hamiltonian
\begin{equation} H'=H+\lambda\sum_{\vec{x}}  \mathcal{G}_{\vec{x}}^\dagger \mathcal{G}_{\vec{x}}.\label{modh}\end{equation}
It is easy to see that \eq{modh} gives an energy of order $\lambda$ to any non-gauge invariant state. As $\lambda\rightarrow\infty$, the non-gauge invariant states decouple from the physical spectrum, but for finite $\lambda$, the gauge-invariant states span the ground state sector of the hamiltonian \eq{modh}, which is gapped. Now, it has been proven that ground states of gapped local hamiltonians satisfy a version of exponential decay of correlations:
\begin{equation} \langle \mathcal{O}_A \mathcal{O}_B\rangle\rightarrow \langle  \mathcal{O}_A P_0 \mathcal{O}_B\rangle=\langle  P_0\mathcal{O}_A P_0 \mathcal{O}_B P_0\rangle,\label{proj}\end{equation}
where $P_0$ is the projector onto the ground state subspace. Equation \eq{proj} tells us that correlators of any operators are identical to correlators of their gauge-invariant parts, as $\lambda\rightarrow\infty$. Thus, we only need to check for exponential decay of correlations in the gauge-invariant theory.

Another problem is that the theorem \eq{alawth} applies to one-dimensional systems, while we have a higher-dimensional conformal field theory. To address this, we can compactify on the spatial manifold $T^{d-2}\times \mathbb{R}$. The transverse torus ensures that we have a  one-dimensional theory, as the theorem requires. In the bulk solution \eq{bbmetric}, this means introducing a periodicity
\begin{equation}x_i\sim x_i+\tilde{L}\end{equation}
 on $(d-2)$ of the transverse coordinates $x_i$. Due to the warping factor, the physical size of the torus then decreases as one goes further into the bulk; to guarantee validity of the Euclidean version of \eq{bbmetric}, one should make sure that the minimal size of the torus is below the cutoff length scale $\Lambda^{-1}$ where the EFT is no longer valid:
 \begin{equation} r_+\,\tilde{L}\gg \Lambda^{-1}.\end{equation}
Combined with \eq{betaf}, one sees that this can be guaranteed by looking at large enough chemical potentials, for any choice of $\tilde{L}$. One worry is that usually in two or three dimensions there can be large quantum effects, which might offset the semiclassical behavior discussed here. A famous case is a massless scalar field, whose propagator falls of with distance in $d>2$ but grows logarithmically in $d=2$. One might ask how this behavior comes about in the dual description. The dominant saddle of a $d$-dimensional CFT on a torus is the so-called $AdS$ soliton - a geometry with a stringy throat in the IR \cite{Horowitz:1998ha}. The two-point function of a scalar in the AdS soliton smoothly interpolates between the UV 2-point function and the IR one as one increases the separation between the operator insertions. From the bulk perspective, this happens because the IR limit is controlled by geodesics which probe the stringy throat of the AdS soliton. 

In the black brane background, however, the geometry caps off before the stringy throat is reached. As a result, geodesics have roughly the same behavior as in the uncompactified setup, and correlators do as well. This is yet another manifestation of the exponential decay of correlations. One may also wonder whether compactification introduces new saddles with lower free energy, but this should not be the case for large enough $\tilde{L}$, where we should recover the noncompact result.

On top of this, \eq{alawth} is derived by a spin chain where the on-site degree of freedom is a single qubit. In the lattice regularization of a holographic field theory, we expect to have a Hilbert space of dimension $D\sim N^2$. \eq{alawth} can be easily restated for a $n-$qubit system via blocking: One writes one site of a $n$-qubit system as $n$ one-qubit sites, and applies \eq{alawth}. Rewriting in terms of $n$-qubit site and blocking, we obtain
\begin{equation}S(L)\leq c\,  n\, l_0 \exp(c' n\,\xi \log(n\xi)).\label{alawth2}\end{equation}
In our case we have a large $N$ theory with $\sim N^2/\epsilon $ degrees of freedom per site (where we have introduced also a cutoff in energies of order $1/\epsilon$), compactified on a $(d-2)$ torus of side $\tilde{L}$. This means we have 
\begin{equation}n\sim\left(\frac{\tilde{L}}{\epsilon}\right)^{d-2} N^2/\epsilon \end{equation}
qubits per site. Here we have taken the lattice cutoff $\epsilon^{-1}$ to also provide the UV cutoff on the Hilbert space of each lattice site. We also need to estimate $l_0$ and $\xi$. The latter is easy: It is simply the correlation length $\mu^{-1}$, in lattice units, so $\xi=\mu\epsilon$. The former is the length scale controlling the onset of the exponential decay of correlations. Since $\mu$ is the only scale in the CFT, we expect $l_0\sim\mu \epsilon$. A more precise way of stating this comes from looking at the mutual information between two regions $A$ and $B$, which controls the decay of correlations \cite{Wolf:2007tdq}:
\begin{equation}I(A:B)\geq \frac{\vert \langle \mathcal{O}_A \mathcal{O}_B\rangle- \langle \mathcal{O}_B\rangle\langle \mathcal{O}_A\rangle\vert}{ \vert\vert \mathcal{O}_{A}\vert \vert \vert\vert \mathcal{O}_{B}\vert \vert}.\end{equation}
Taking $A$ and $B$ to be two strips in a black brane background, the mutual information vanishes after a critical separation of order $(r_+/\ell)^{-1}\sim \mu^{-1}$ \cite{Fischler:2012uv,Andrade:2013rra}. From this point on, $I(A:B)\lesssim\mathcal{O}(1)$, so correlations are small. It is then reasonable to estimate  $l_0\sim\xi$. One could also compute $l_0$ explicitly, by just computing the correlators of single-trace operators. 

Putting all of this together, we have
\begin{equation}S(L)\leq c  p  \exp(c' p\log p),\quad p\equiv\left(\frac{\tilde{L}}{\epsilon}\right)^{d-2} \frac{N^2\mu}{\epsilon^2}.\label{alawth3}\end{equation} 
This is the version of the area law one can hope to have in a large $N$ holographic CFT, with the caveats mentioned above. It has the expected prefactor of $N^2$ and dependence on the cutoff. As discussed in Section \ref{sec:EE}, only the universal part of the entanglement entropy, which is cutoff-independent, has a physical meaning independent of the lattice regularization. For this reason, it is convenient to recast the result \eq{alawth3} in terms of a mutual information, which is insensitive to the details of the regularization.  In our case we have a thermofield double state, and the bipartition consists of the union $A_L\cup A_R$  of two identical strips of length $L$ on both sides of the CFT, and their complement. The mutual information
\begin{align} I_{\ket{\Psi}}(A_L:A_R)&=S_{\ket{\Psi}}(A_L)+S_{\ket{\Psi}}(A_R)-S_{\ket{\Psi}}(A_L\cup A_R)\nonumber\\&=2S_{\ket{\Psi}}(A_L)-S_{\ket{\Psi}}(A_L\cup A_R)\nonumber\\&=2(S_{\ket{\Psi}}(A_L)-S_{\ket{0}}(A_L))- (S_{\ket{\Psi}}(A_L\cup A_R)-2S_{\ket{0}}(A_L))\label{mitfd}\end{align}
is a purely IR quantity insensitive to the UV regulator. Now take the strip size $L$ to be very large, and we have added and substracted the entanglement entropy of the CFT vacuum (without chemical potential). In the limit $L\rightarrow\infty$, \eq{alawth3} guarantees that the last term in \eq{mitfd} stays bounded, so that the asymptotic behavior of $I_{\ket{\Psi}}(A_L:A_R)$ is controlled by the entanglement entropy difference $S_{\ket{\Psi}}(A_L)-S_{\ket{0}}(A_L)$. Entropy differences are also UV insensitive, so this expression is meaningful. In other words, if $S_{\ket{\Psi}}(A_L)-S_{\ket{0}}(A_L)$ scales faster than the area of $A_L$ (and this will be the case for us, since the one-sided entropy scales like a volume, see \eq{lala}) , we have
\begin{equation}\lim_{L\rightarrow\infty} \frac{I(A_L:A_R)}{S_{\ket{\Psi}}(A_L)-S_{\ket{0}}(A_L)}\rightarrow2.\label{alawth4}\end{equation}

Equations \eq{alawth3} and \eq{alawth4} are the core results that we want to compare to the black brane background. But before we do so, let us mention a couple of problematic points:
\begin{itemize}
\item Theorem \eq{alawth} requires exponential decay of \emph{all} localized correlators. This includes not only local operators, but also operators built out of arbitrarily many insertions, as long as these are done in a finite region. In the large $N$ limit, and for single-trace operators, one can extend the result for the two-point function to local operators built of up to $N$ single-trace operators, via large $N$ factorization, but this technique does not work forever \cite{ElShowk:2011ag}. At some point, gravitational backreaction becomes strong enough and large $N$ factorization does not work.  In our setup this corresponds to very massive black hole microstates, with a small charge-to-mass ratio. I do not know how to prove that correlators of these vanish exponentially, and can only offer some plausibility arguments. In a correlator of the form
\begin{equation} \langle \Psi \vert A_X B_Y\ket{\Psi}\end{equation}
where $A_X$ is an arbitrary operator supported in a region $X$, and $B_Y$ is another such operator supported in a region $Y$, one can in principle OPE all of the local operators constructing $A_X$, and the same in $Y$, to obtain an expression of the form
\begin{equation} \sum_{\mathcal{O},\mathcal{O}'}\int_X dx \int_Y dy\, K_{A_X,\mathcal{O}(x)}\, K'_{B_Y,\mathcal{O}'(y)} \langle \mathcal{O}(x) \mathcal{O}'(y)\rangle_{\ket{\Psi}},\end{equation}
where $K_{A_X,\mathcal{O}(x)}$ and $K'_{B_Y,\mathcal{O}'(y)}$ are kernels obtained by explicitly carrying out the OPE\footnote{Since the ground state is not invariant under dilatations, we must understand the OPE as an asymptotic expansion valid when the separation is small, and not a convergent series as would happen for correlators in the CFT vacuum.} in both $A_X$ and $B_Y$. One expects two-point functions to vanish exponentially even for operators of high dimension, since this is suggested by the worldline approximation \cite{Andrade:2013rra}. 

Later, in Subsection \ref{sec:wgc}, we will see that things improve a little for near-extremal, small black holes (compared to the AdS radius). Here, the usual balance of electric and gravitational forces leads to low near-horizon scaling dimensions, and interactions effectively ``switch off''. While these are not single-trace operators in the usual sense of the word, they can be described semiclassically, and the fact that their interactions switch off suggests that we recover some notion of factorization. However, this argument fails for operators dual to large black holes (in AdS units). 

\item Relatedly, the upper bound \eq{alawth3} grows very fast with $N$, so one is precluded from taking the strict $N\rightarrow\infty$ limit. Rather, we must work at very large but finite $N$, and then look at the entanglement entropy of very large strips, whose length grows exponentially with $N$. As emphasized in the Introduction, and also in Section \ref{sec:pre}, we will be assuming that even for these very large strips the leading $N$ piece gives the leading contribution to the entanglement entropy. In other words, effective field theory is reliable. Similarly, requiring $\Delta(\vec{p})$ to be holomorphic on a strip around the real axis is a statement which is sensitive to $1/N$ corrections; these might in principle might in principle shift or introduce additional poles and/or branch cuts. In Section \ref{sec:wgc}, we will apply the above argument using large $N$ expressions for $\Delta(\vec{p})$; physically, this amounts to saying that the (weak) gravitational interactions do not somehow result in more extremal bound states (as measured by the near-horizon effectivemass). 
\end{itemize}

With these caveats in mind, the bottomline is that \eq{alawth} tells us that the entanglement entropy in the $T=0$ thermofield double state should satisfy an area law \eq{alawth3}, or in terms of UV-insensitive quantities, \eq{alawth4}. Let's check if the bulk agrees. The holographic bulk computation of the relevant entanglement entropy was analyzed in \cite{Andrade:2013rra}. A similar analysis for the eternal Schwarzschild black hole was carried out in \cite{Hartman:2013qma}. 

\begin{figure}[!htb]
\begin{center}
\includegraphics[width=9cm]{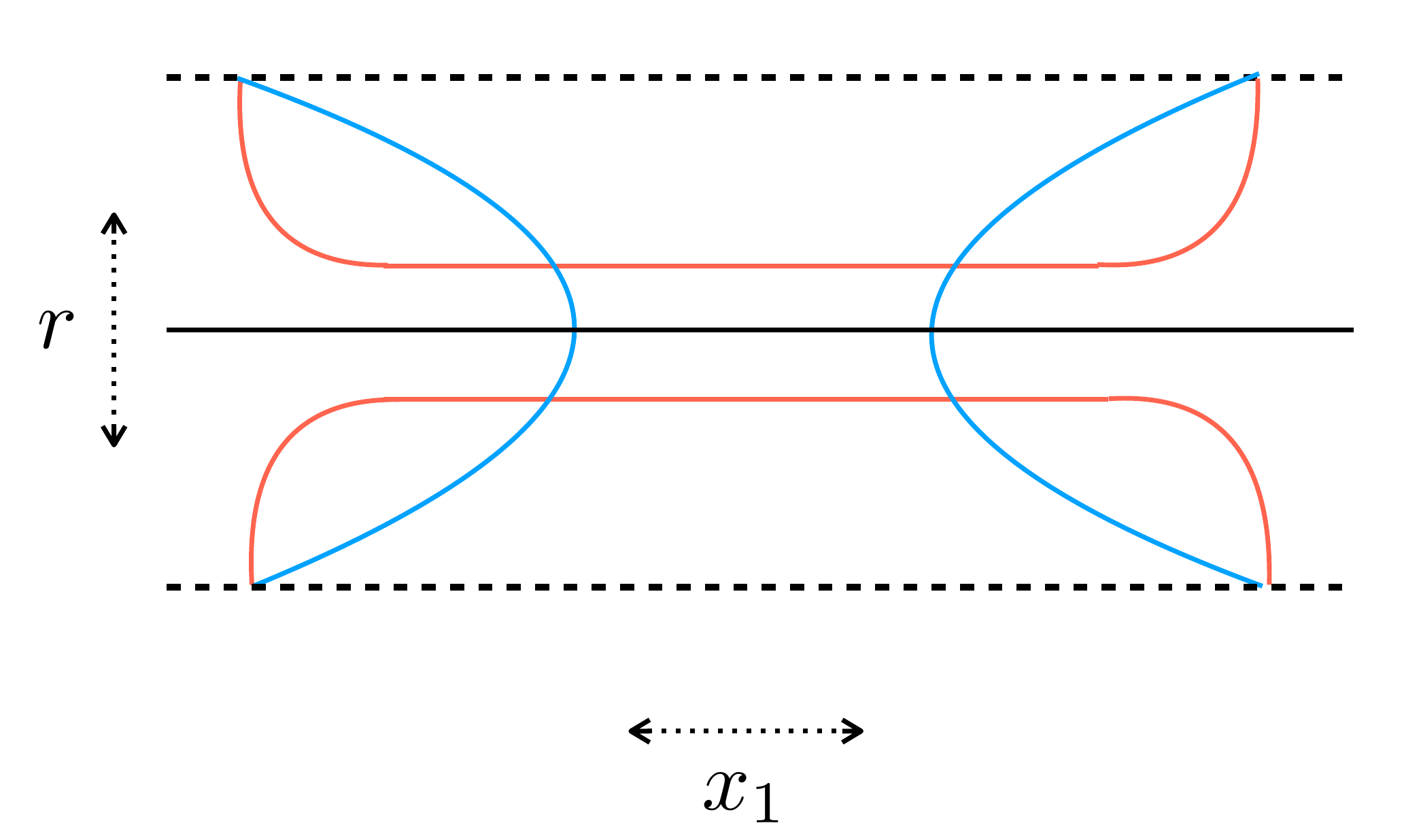}
\end{center}
\caption{Illustration of the different entangling surfaces in the thermofield double. The whole setup is taken at $t=0$. The dotted horizontal lines represent the two boundaries, and the solid black line in the middle represents the black brane horizon. The entangling surface consists of two identical pieces, one in each boundary. The red entangling surface is similar to two copies of what one would have in the one-sided black hole, while the blue one goes through the horizon. }
\label{f3}
\end{figure}

We need to compute the entanglement entropy of a strip of length $L$ in the dual theory. We should regard the two uncoupled CFT's as different sectors of the same local field theory, which means that the Ryu-Takayanagi surfaces must anchor at two identical strips on each side of the eternal black hole. The situation is depicted in Figure \ref{f3}. As we can see, there are two extremal surfaces satisfying the homology constraint (this requires the RT surface to be equivalent in singular homology to the boundary entangling region \cite{Rangamani:2016dms}) . One of them can be contracted to the boundary, and the other goes through the wormhole. 

The whole point of this discussion is that the two different extremal surfaces lead to different behavior for the boundary entanglement entropy. The contractible surface, which does not go through the horizon, is just two copies of the one-sided Ryu-Takayanagi surface, already discussed in Section \ref{sec:pre}. It therefore leads to a volume law for the entanglement entropy. The non-contractible surface, on the other hand, leads to (see Appendix \ref{app:C}) \cite{Andrade:2013rra}
\begin{align}S^{\text{n.c}}_{\ket{\Psi}}(L)\approx\frac{2\pi r_+^{d-1}}{\kappa_{d+1}}V_{d-2}\frac{\log(\beta \Phi/\ell)}{\sqrt{2}(d-2)\Phi}.\label{ncon}\end{align}
Here, $V_{d-2}$ is the volume of the transverse (d-2)-torus. Notice that there is no $L$-dependence in \eq{ncon}, so it is describing an area law for the entanglement entropy.

When there is more than one extremal surface satisfying the homology constraint, one is always supposed to take the one with the smallest (bulk) area. Since the non-contractible result \eq{ncon} diverges logarithmically as $T\rightarrow0$, the Ryu-Takayanagi prescription predicts a volume law for the entanglement entropy of the ground state in the boundary field theory\footnote{Note that we obtain the $T\rightarrow0$ results as limits of finite $T$ ones. The idea is that one first fixes $L$ and then sends $T\rightarrow0$. In this way we do obtain a volume law for any $L$.}. This was already noticed by the authors of \cite{Andrade:2013rra}, which describe a vanishing mutual information in the extremal limit.

Since according to \eq{alawth} and \eq{alawth3}, only an area law should be admissible, we arrive at the advertised contradiction and conclude that the theory must somehow be inconsistent, so that the above analysis breaks down. We also see a violation of \eq{alawth4}; The denominator grows like $N^2$, while the numerator is $\mathcal{O}(1)$.

We can pinpoint precisely why the pathological behavior shows up: It is because of the logarithmic divergence in \eq{ncon}. Notice that in fact an area law is predicted at any $T\neq0$, precisely because the finiteness of \eq{ncon} ensures that for large enough $L$ the contractible surface will have a larger area. The same happens for the Schwarzschild black hole; in fact it was noticed in \cite{Hartman:2013qma} that the thermofield double behaves as a gapped state, with exponentially small correlations and area law entanglement.

In turn, the logarithmic divergence in \eq{ncon} boils down to the fact that the extremal wormhole throat becomes infinite, effectively disconnecting the two CFT's. More concretely, the area element of the extremal RT surface goes as 
\begin{equation}dA\sim  V_{d-2} \frac{r^{d-2}\,dr}{\sqrt{U(r)}}.\end{equation}
Near $r\approx r_+$, the $r^{d-2}$ factor is approximately constant, and we obtain a logarithmic divergence,
\begin{equation}dA\approx  V_{d-2} \frac{r_+^{d-2}}{\ell_{AdS_2}}\frac{dr}{r}.\label{logdiv}\end{equation}
This divergence is the same one obtains in a radial infalling geodesic.

\begin{figure}[!htb]
\begin{center}
\includegraphics[width=7.5cm]{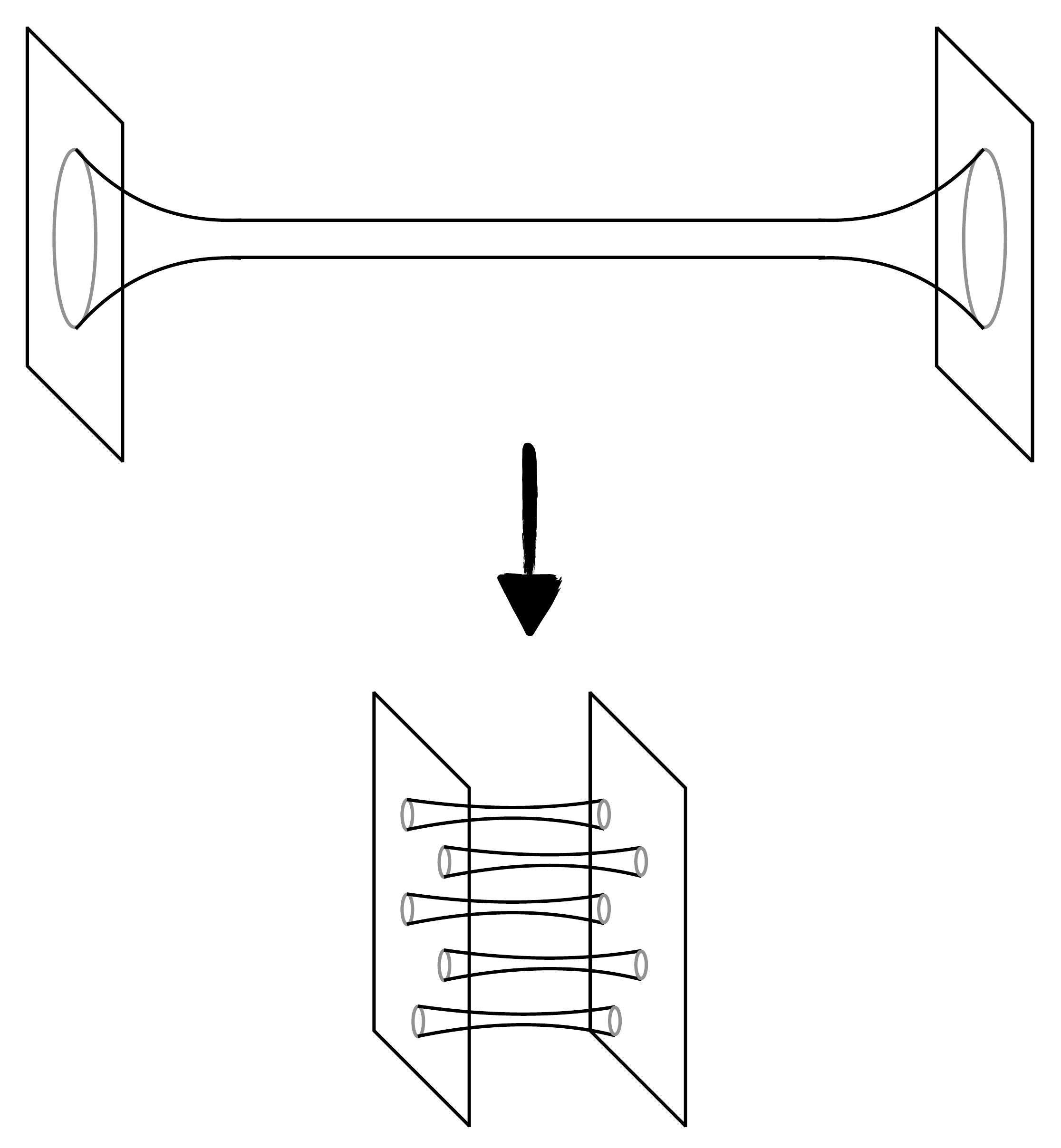}
\end{center}
\caption{On top we have the (near-extremal) black brane, connected by a very long wormhole whose length diverges in the extremal limit. The WGC triggers an instability and the black brane decays to a myriad of ``Planck-sized wormholes'', which are presumably OK with ER=EPR.}
\label{fb1}
\end{figure}

The result is reminiscent of the ER=EPR proposal \cite{Maldacena:2013xja}, where entanglement is somehow dual to the existence of a bulk wormhole geometry. The proposal essentially says that if we have some state with a horizon in the bulk, and purify it via the thermofield construction, the bulk solution should develop a wormhole connecting the two sides. However, in the extremal black brane, the throat becomes infinitely long, and the wormhole effectively disconnects. So we get a contradiction with \eq{alawth3} precisely at the time where ER=EPR seems to fail.  This failure led \cite{Maldacena:2013xja} to claim that a better description of the thermofield double \eq{tfd} is actually an almost extremal black hole (such that the temperature is below the gap of the system), for which the wormhole throat is finite. While this works for black holes, there is no gap for an infinite black brane, so maybe in this case the absence of a wormhole is a real pathology that needs to be cured. And this can happen if one introduces light enough fields, which cause the wormhole to branch out into smaller wormholes until the semiclassical picture no longer applies. In the supersymmetric case, this branching out was described in \cite{Maldacena:1998uz}; in the nonsupersymmetric case, where the black brane decays, the wormhole branches out into a myriad Planckian wormholes connecting the elementary particles in each AdS factor, as depicted in Figure \ref{fb1}.

The result \eq{alawth4} can also be obtained using the long-distance expansion for the mutual information in a QFT in terms of local operators \cite{Cardy:2013nua,Faulkner:2013ana}. This has the advantage of working directly in $d$ dimensions. For a thermofield double, we obtain that the low-temperature limit of the mutual information is governed by a two-point function
\begin{equation} I(A_L:A_R)\sim \left( \langle \mathcal{O}_L(\vec{0},0) \mathcal{O}_R (\vec{0},0)\rangle - \langle\mathcal{O}\rangle^2\right)^2,\label{minfexp}\end{equation}
where $\mathcal{O}$ is the operator with slowest-decaying two-sided correlator. Generically, this vanishes as $\beta^{-2\Delta_{\mathcal{O}}}$ \cite{Faulkner:2013ana}, so if the operator of lowest dimension has $\Delta_{\mathcal{O}}>0$, then the mutual information must vanish in the extremal limit, in contrast with the holographic result. Notice that the assumption of holomorphicity for $\Delta(\vec{p})$ (which ensures exponential decay of correlations) was not needed in this case; the result obtained is thus slightly stronger. Another advantage is that \eq{minfexp} can be understood as a result purely in the large $N$ limit; one does not need to work at finite $N$.

\subsection{A nonperturbative formulation of the WGC}\label{sec:wgc} 
After going through the argument, we are finally in a position to write down the main claim of this paper:

\begin{center}\bluebox{\textbf{In any consistent quantum theory of gravity in AdS, there can be no extremal black branes with a near-horizon $AdS_2$ geometry in which the spectrum of $iD$ is holomorphic on an open strip around the real axis and gapped.}}\end{center}

The claim is that this is a precise, nonperturbative reformulation of the WGC in the $AdS$ context. The reader is perhaps annoyed at this point: The result looks nothing like the WGC, which is a precise inequality involving numbers, rather than just words. To see the relationship between the two, it is best to focus on an example. We will go back to our benchmark, the Einstein-Maxwell theory with the black brane \eq{bbmetric}, and introduce elementary particles of charges $q_i,m_i$ and spins up to 1. These are coupled in the standard way to the gauge field and the metric.  The very well known result \cite{Hartnoll:2016apf} is that the near-horizon spectrum is of the form

\begin{equation}\Delta(\vec{p})=\frac12+n+\sqrt{\frac14+\frac{\ell^2}{d(d-1)}\left(m^2-\frac{g^2}{\kappa_{d+1}^2}q^2+\frac{d(d-1)}{(d-2)^2}\frac{g^2\vert\vec{p}\vert^2}{\kappa_{d+1}^2\mu^2}\right)}\label{massy}\end{equation}
for scalars and
\begin{equation}\Delta(\vec{p})=\frac12+n+\sqrt{\frac{\ell^2}{d(d-1)}\left(m^2-\frac{g^2}{\kappa_{d+1}^2}q^2+\frac{d(d-1)}{(d-2)^2}\frac{g^2\vert\vec{p}\vert^2}{\kappa_{d+1}^2\mu^2}\right)}\label{massy2}\end{equation}
for fermions. We should worry also about the metric and gauge field; the near-horizon spectrum associated to these was computed in \cite{Edalati:2010hk} (see also \cite{Hartnoll:2016apf}), and is both holomorphic in the momentum in a neighbourhood of zero and lower-bounded by $1/2$. 

We should also take into account charged string states, black holes, and anything that might be in the UV. In fact, there is increasing evidence in the flat space setup that the WGC is satisfied by UV objects, such as black holes \cite{Hamada:2018dde}. We will deal with these heavy objects via a mean field theory calculation, in which one neglects the backreaction of the objects on the metric. In this approximation, the motion a heavy object of mass $M$ and charge $Q$ is described via an effective quantum-mechanical model \cite{Pioline:2005pf}:
\begin{align}&\left(\Delta-\frac{\mathcal{E}}{\tan^2\sigma}\right)^2\Psi(\sigma)=\left(-\frac{d^2}{d\sigma^2}+\frac{\mathcal{M}^2}{\sin^2\sigma}\right) \Psi(\sigma),\nonumber\\\mathcal{M}^2&\equiv \frac{\ell^2}{d(d-1)}\left(m^2+\frac{d(d-1)}{(d-2)^2}\frac{g^2\vert\vec{p}\vert^2}{\kappa_{d+1}^2\mu^2}\right),\quad \mathcal{E}\equiv \frac{\ell}{\sqrt{d(d-1)}}\frac{gQ}{\kappa_{d+1}}.\label{qmmodel}\end{align}
In the limit of large conformal dimension, the results can be well approximated by the Born-Sommerfeld quantization rule, leading again to \eq{massy}. When the conformal dimension is lower than one, there is an alternate boundary condition, as usual in AdS/CFT.  Which choice of the square root one takes depends on the boundary conditions of the problem. These are in turn inherited from the boundary conditions of the full AdS solution. In any case, we should note that this ambiguity only shows up when $\mathcal{M}\approx\mathcal{E}$, i.e. for states that are (almost) extremal. 

Physically, the main point of the above analysis is that the near-horizon region of a large black brane just looks like AdS space itself with a constant electric field turned on. We have just worked out the dynamics of the different charged objects of the theory in such a background, assuming they are stable (we neglected interactions). Presumably, some of the more heavy objects will actually decay into lighter ones  (just like a neutral atom ionizes when subjected to a strong enough electric field), but the above analysis should be valid more generally for the lightest charged objects in the theory, for each value of the charge, whenever the $U(1)$ coupling is small.

In any case, the above analysis shows that in the weakly coupled regime, the near-horizon conformal dimensions are directly related to masses and charges of the states of the theory.  The only thing left to check is holomorphicity of the spectrum of $iD$ near the real axis. Looking at \eq{massy}, we see that for bosons this is indeed the case if and only if the near-horizon effective dimension is above the $BF$ bound: 
\begin{align}m^2-\frac{g}{\kappa_{d+1}}q_{\phi}^2\leq \frac{d(d-1)}{4\ell^2}.\label{adswgcfull}\end{align}
This is one of the versions of the WGC introduced by Nakayama and Nomura in \cite{Nakayama:2015hga}. Introducing the charge-to-mass ratios of the extremal black hole\footnote{This is the charge-to-mass ratio of a small extremal black hole (compared to the $AdS_{d+1}$ radius) or equivalently an extremal asymptotically flat RN black hole. It is therefore the charge-to-mass ratio that shows up in traditional WGC arguments.} and particle respectively,
\begin{align}\xi\equiv\frac{q_{\phi}}{m},\quad \xi_{\text{ext.}}=\frac{c^2}{2}\frac{\kappa_{d+1}}{g},\end{align}
\eq{adswgcfull} can be rewritten as
\begin{align}\frac{\xi}{\xi_{\text{ext.}}}>\left(\frac{d-1}{d-2}\right) \left(1+\frac{d(d-1)}{4(m\,\ell)^2}\right).\label{adswgcfull1}\end{align}

Equations \eq{adswgcfull},\eq{adswgcfull1} have an interesting consequence: As already pointed out in \cite{Nakayama:2015hga}, the bound  \eq{adswgcfull1} is stronger than the usual WGC bound derived from flat space heuristic arguments, which is simply
\begin{align}\xi>\xi_{\text{ext.}},\end{align}
by an $\mathcal{O}(1)$ factor. This is an intrinsically $AdS$ effect which has to do with the fact that in large extremal RN-AdS black holes the near-horizon $AdS_2$ radius is locked to the asymptotic $AdS_{d+1}$ radius, while no such phenomenon occurs for flat space black holes. However, this only happens if one looks at bosons; one can see from \eq{massy2} that the same effect does not take place for fermions.

As mentioned above, the last term in the bound \eq{adswgcfull1} is related to a particular boundary condition in the near-horizon geometry of \eq{bbmetric}. In a more general but still weakly coupled theory this boundary condition might change; we will see this is indeed the case in supersymmetric theories. There is a potential issue with  boundary conditions whenever 
\begin{equation}\mathcal{M}\leq \mathcal{E}.\label{sssa}\end{equation}
 For a weakly coupled theory with states satisfying \eq{sssa} figuring out whether it complies with the statement above requires extra work to understand the boundary conditions. But we can immediately exclude any theory in which \emph{all} states violate \eq{sssa}. This particular case of the statement above is sufficiently interesting that it deserves another blue box:

\begin{center}\bluebox{\textbf{(The Weak Gravity Conjecture in AdS)} In any consistent weakly coupled quantum theory of gravity weakly with exactly stable charged extremal black branes with a near-horizon $AdS_2\times\mathbb{R}^{d-1}$ geometry, there has to be at least one state which satisfies
\begin{equation}m^2\leq \frac{g}{\kappa_{d+1}}q_{\phi}^2.\label{wgc}\end{equation}
}\end{center}

We should emphasize that \eq{adswgcfull} or \eq{wgc} are not simply BPS bounds in disguise. Even in supersymmetric theories, one can have $U(1)$'s which do not show up as central charges in the supersymmetry algebra. In this case, there is no BPS bound, even if the theory and the vacuum under consideration are supersymmetric. The 10d right-moving heterotic $U(1)$ originally used as an example of the WGC is precisely of this kind \cite{ArkaniHamed:2006dz}. In AdS string compactifications in Sasaki-Einstein manifolds, these are typically flavor $U(1)$'s, which in the bulk arise from dimensional reduction of $p$-form gauge fields on nontrivial cycles of the compactification manifold \cite{Butti:2005sw}.

However, it is true that, whenever we discuss a supersymmetric black brane, \eq{wgc} becomes the BPS bound. We know examples of BPS asymptotically AdS black branes, for instance in gauged supergravity \cite{Hristov:2012bd}, which preserve $1/4$ of the supercharges. Clearly, these black branes must be exactly stable. As a result, the spectrum of $iD$ is automatically real and positive. To be consistent with \eq{wgc}, $iD$ must either be nonlocal, as in the electric examples above, or it may also be gapless. In supersymmetric theories, this is often the case because one has BPS local operators. To be precise, since the near-horizon geometry is 1/4 BPS, the conformal algebra \eq{so21algebra} is enhanced to a 1d superconformal algebra. This involves the introduction of two near-horizon supercharges $Q,S$, which satisfy \cite{BrittoPacumio:1999ax}
\begin{equation}\{Q,S\}=D.\end{equation}

Consider a BPS local operator $\mathcal{O}(\vec{x},t)$ which preserves the same supercharges as the background black brane. Just as with the scalar field example, we can expect that $\mathcal{O}(\vec{x},t)$ will decompose in KK modes, leading to a continuous spectrum $\Delta(\vec{p})$. The zero mode of the BPS primary will be annihilated by the supercharge $\mathcal{Q}$, and thus will have $\Delta=0$, so we have a gapless spectrum. Physically, the BPS operator adds one additional charged particle to the black brane, increasing its charge by an infinitesimal amount, leading to a slightly different black brane, with a higher charge, but still with a transparent horizon. One can also understand this in the mean field model \eq{qmmodel}: supersymmetry forces us to take the negative square root, leading to $\Delta=0$.

A couple of final points. First, in contrast with every other version of the WGC known to me, the main statement is nonperturbative in any EFT coupling, i.e. it involves only considerations about the spectrum of physical excitations over the extremal black hole background. By contrast, the standard formulation of the WGC as in \eq{wgc} involves explicitly the $U(1)$ gauge coupling $g$, which is assumed to be small (the statement above still requires Einsteinian gravity). Second, the statement is UV sensitive, since in principle even very massive fields or black holes can in principle lead to a gapless or complex spectrum; in this sense, the above statement includes the possibility that the WGC is satisfied via black holes. 
 
Finally, there might be other ways to obtain a gapless spectrum not directly related to supersymmetry -- for instance, one could imagine an infinite tower of states whose near-horizon dimensions scale as $1/n$ --. I do not know of any examples where something like this is realized, but if possible, they would perhaps correspond to exotic ways to comply with the WGC. We will speculate about a system that might realize this behavior in the Conclusions.

\section{Discussion and conclusions}\label{conclus} 
It is often said that consistent quantum theories of gravity must satisfy the WGC in order to avoid stable extremal black holes, but it was never clear precisely what is wrong with these. The best one could do in this line was heuristic arguments involving remnants in a limit ($g\rightarrow0$ at finite $M_P$) which does not necessarily exist.  

This note helps to fill in that gap in the AdS context, by pointing out that extremal exactly stable black brane solutions are actually inconsistent with quantum information theorems. More concretely, the volume law predicted by the Ryu-Takayanagi formula means that the spectrum of the near-horizon dilatation operator cannot be both holomorphic near the real axis and gapped. Physically, this means that extremal black branes must either be (on the verge of) an instability, or have a zero mode. The former case correspond to the standard WGC, since the near-horizon scaling dimensions are related to masses and charges of UV states. The latter covers the supersymmetric case. 

 This property, which must hold in any consistent quantum theory of gravity, may be regarded as a nonperturbative generalization of the WGC, valid even at strong (gauge) coupling. At this point it is probably good to summarize the assumptions involved, which are:
 \begin{enumerate}
 \item That the two-sided black brane solution (compactified to $2+1$ dimensions) is dual to a thermofield state in the dual field theory.
 \item Validity of the large $N$ approximation, i.e. that the semiclassical picture gives the leading piece of the quantities of interest (in particular, the near-horizon $AdS_2$ geometry controlling the correlators and the Ryu-Takayanagi prescription for entanglement entropy).
 \item  That the holographic dual CFT admits a lattice or lattice gauge theory formulation.
  \item That the semi-local critical behavior seen for single-trace operators of low dimension also works for $n$-point functions and for operators of high dimension. 
 \end{enumerate}
 The last point is the weakest link in the chain. It is hard to argue on general grounds that the semi-local behavior holds for primaries of large dimension, beyond the reach of large $N$ factorization, although this is suggested by the worldline approximation. Also, high-dimension operators belong in a handwavy sense to the UV of the theory; one expects that IR quantities such as the long distance behavior of the entanglement entropy or mutual information are not very sensitive to them.  In this sense, the second point can be recast as the requirement that the IR properties of the entanglement entropy or mutual information are not UV sensitive, which has been argued before \cite{Casini:2015woa}. It is reassuring that an argument based on a long-distance expansion of the mutual information in terms of two-point functions leads to the same result, directly in $d$ dimensions and requiring only $\Delta_{\mathcal{O}}>0$.
  
 Other than that, the argument given here is very general - it only relies on having Einsteinian gravity, and an extremal bulk black brane solution. In practice, any holographic model that predicts a (degenerate) ground state with a finite area horizon is in trouble.  On the other hand, this generality means that we only have an argument for the mildest version of the WGC - it would be interesting to see how far can we get on the road towards e.g. a Lattice WGC \cite{Heidenreich:2015nta,Heidenreich:2016aqi,Montero:2016tif,Andriolo:2018lvp}. Such an improvement is necessary if one wants to provide meaningful constraints to EFT's. 

The result obtained here also has some connection to the ER=EPR correspondence. The pathological behavior of the black brane solution is directly connected to the fact that the length of the wormhole throat in the two-sided extremal black brane solution is actually infinite, so that the two sides of the black brane effectively disconnect. According to ER=EPR this would mean no entanglement, yet one can see explicitly from \eq{tfdt0} that the two sides are actually heavily entangled. Thus ER=EPR suggests that the two sides cannot have really disconnected, and so the black brane picture must be corrected. The result derived here does precisely this, by either introducing long-range correlations that can traverse the wormhole or by introducing an instability, which breaks it up into a myriad Planckian wormholes.

The Einstein-Maxwell effective field theory in AdS is quite popular in bottom up holography and AdS/CMT. Based on the above discussion, can one conclude that results obtained with this model are inconsistent? Not necessarily; the inconsistency may well be harmless depending on the particular question one is trying to answer. As an analogy, consider a four-dimensional $SU(2)$ gauge theory with one fermion in the fundamental. This theory is anomalous \cite{Witten:1982fp} and therefore it does not really make sense. However, due to the nonperturbative character of the anomaly, as long as we only consider perturbative quantities the theory seems to work. On the other hand there are observables such as e.g. the partition function which are pathological. Similarly, perhaps computing one or two-point functions in Einstein-Maxwell-AdS yields reasonable results that make sense as models of condensed matter systems; on the other hand, we have just seen that one should not use this model at all to make statements about long-range entanglement.

While in this note we have focused on the case of an ordinary $U(1)$ gauge symmetry, the argument works in the same way for gauged $p$-form generalized global symmetries, in the range $0\leq p \leq (d-1)$, by considering higher codimension branes. The ``$0$-form gauge field'' (axion) and $d$-dimensional gauge fields (membranes) remain outside of the scope of this analysis, although it would be interesting to study them from the entanglement entropy perspective. The $d$-form case is particularly relevant, as it has to do with stability of AdS space \cite{Ooguri:2016pdq}.

Can the above arguments be extended to asymptotics other than AdS? Clearly not in a rigorous way, at least at present. We lack enough information about the putative holographic duals to apply \eq{alawth}. I also do not know of a rigorous way to compute entanglement in spacetimes with arbitrary asymptotics (see however \cite{Solodukhin:2011gn}). However, the statement that a local gapped near-horizon dilatation operator\footnote{Understood in the usual condensed matter sense of a gap in the infinite volume limit. Near-horizon geometries of spherical black holes have a gap below which the semiclassical description breaks down \cite{Almheiri:2016fws}. This gap decreases as a power of the black hole radius, so it is not an issue in the infinite volume (large charge) limit.} is problematic makes sense in dS and Minkowski as well, so maybe theories where this happens should be regarded with suspicion. 

An example of a case that is not readily accesible with AdS asymptotics is that of extremal rotating black holes, which also have a near-horizon $AdS_2$ geometry. While the Kerr-AdS metric does exist \cite{Carter:1968ks}, it does not admit a scaling limit leading to a black brane. So let us look at an asymptotically flat extremal Kerr black hole instead, and consider for simplicity a scalar field of any mass $m$. The near-horizon spectrum of perturbations approaches extremality asymptotically \cite{Hod:2016iri}, so the near-horizon spectrum is gapless even for large black holes. Thus we get no Swampland constraints from rotating black holes, which was to be expected from the heuristic point of view, since a rotating black hole can always shed some of its angular momentum by emitting elementary particles in high angular momentum configurations.

The big question addressed in this paper is  ``What's wrong with a WGC-violating EFT?'' The derivation presented here gives an answer -- entanglement --. So perhaps it is a good idea to further look more closely at entanglement to get more Swampland constraints! This is, in my opinion, the main appeal of providing a solid argument for a Swampland conjecture - it tells us where to look for even more such constraints.

In this line, it is possible that other familiar Swampland conjectures, such as part of the Swampland Distance Conjecture \cite{Ooguri:2006in} or the Scalar Weak Gravity Conjecture \cite{Palti:2017elp,Lust:2017wrl} can be understood via entanglement. I hope to return to this point in the near future.

A particular model where the results of this paper could be studied more precisely the SYK chain \cite{Gu:2016oyy}. This is a (1+1)-dimensional lattice model with semi-local criticality and a finite ground state entropy. Theorem \eq{alawth} would then naively imply an area law for the thermofield double entanglement entropy, for arbitrarily low temperatures\footnote{Although one would have to revisit the hypothesis of the theorem due to the averaging over couplings involved in SYK.}. The way the model avoids the contradiction discussed in the main text could be related to the fact that the gravitational dual of SYK is not expected to be Einstein \cite{Sarosi:2017ykf}, so one should not compute entanglement entropies via the RT prescription.

\subsection*{Acknowledgements}
It is a pleasure to thank Angel Uranga, Anthony Charles, Dionysos Anninos, Eduardo Martin-Martinez, Gary Shiu, I\~{n}aki Garcia-Etxebarria, J. L. F. Barb\'{o}n, Javier Martin-Garcia, John Stout, Juan Pedraza, Luis Iba\~{n}ez, Nabil Iqbal, Pablo Bueno, Ramis Movassagh, Takaaki Ishii, Thomas Grimm, Thomas Van Riet, Toshifumi Noumi and especially Markus Dierigl, Kilian Mayer, Irene Valenzuela, Wilke van der Schee, and Gianluca Zoccarato for very useful discussions and comments. I also thank Perimeter Institute, the University of Wisconsin-Madison, and Cornell University for hospitality while this work was being finished. I am supported by a Postdoctoral Fellowship of the Research Foundation - Flanders, although a significant part of this work was completed when I was supported by a Postdoctoral Fellowship from ITF, Utrecht University.

\appendix

\section{Computation of entanglement entropies}\label{app:C}
In this Appendix we will compute the results \eq{lala} and \eq{ncon} of the main text, using the Ryu-Takayanagi formula. We will anchor the RT surfaces at radius $r=1/\epsilon$, related to the lattice spacing in the dual field theory. For a more detailed account of these results (for $d=4$) see \cite{Rangamani:2016dms,Andrade:2013rra,Kundu:2016dyk}.

\subsection{Entanglement entropy for rectangular regions}
We will take the boundary entangling surface to be a rectangle in $\mathbb{R}^d$, with $(d-2)$ sides of lengths $L_i$, $i=2,\ldots (d-1)$ with cross section $V_{d-2}$, and a much longer one along $x_1$ of length $L\gg L_i$. We will parametrize the surface by the coordinate $r(x_1)$, with the midpoint at $x_1=0$. The area is then given by
\begin{align}\mathcal{A}=V_{d-2}\int_{-L/2}^{L/2} r^{d-2} \sqrt{\dot{r}^2/U(r)+r^2}\,dx_1,\label{areafu}\end{align}
with boundary conditions $r(-L/2)=r(L/2)=1/\epsilon$. Invariance under $x_1$ translations means there is a conserved quantity
\begin{align}E\equiv \frac{r^{d}}{ \sqrt{\dot{r}^2/U(r)+r^2}}=r_t^{d-1},\label{e0}\end{align}
where $r_t$ is the radius at the turnaround point. Equation \eq{e0} can be integrated to yield
\begin{align}\frac{L}{2}&=\int_{r_t}^{1/\epsilon} \frac{dr}{\sqrt{U(r)\left(\frac{r^{2d}}{r_t^{2d-2}}-r^2\right)}},\label{ee11}\\\mathcal{A}&=2V_{d-2} \int_{r_t}^{1/\epsilon} \frac{r^{2d-2}}{r_t^{d-1}}\frac{dr}{\sqrt{U(r)\left(\frac{r^{2d}}{r_t^{2d-2}}-r^2\right)}}.\label{ee22}\end{align}
In the $L\rightarrow\infty$ limit, $r_t\approx r_+$, and the extremal curve hovers just above the horizon for most of its length. In this case, one can plug \eq{ee11} in \eq{ee22} to get the estimate
\begin{equation}\frac{2\pi \mathcal{A}}{\kappa_{d+1}} \approx V_{d-2}\,L \frac{2\pi r_+^{d-1}}{\kappa_{d+1}},\end{equation}
which is precisely \eq{lala}. 

We are also interested in the entanglement entropy in the thermofield double state, where the entangling surface is two copies of the rectangle described above, one in each boundary. In the Euclidean picture, the second boundary can be taken to lie at Euclidean time $t_E=\beta/2$ \cite{Andrade:2013rra}. We will parametrize the curve by $x_1(r)$. Looking at \eq{areafu}, it is then clear that $x_1=\text{const.}$ is the extremal solution. One then gets directly
\begin{align}\frac{4\pi \mathcal{A}}{\kappa_{d+1}}=\frac{2\pi}{\kappa_{d+1}}V_{d-2}\int_{r_+}^{1/\epsilon}  \frac{r^{d-2}}{\sqrt{U(r)}}d\,r.\label{areafu2}\end{align}
This expression works also away from extremality. There is no $L$ dependence, so this is describing an area law. The near-horizon contribution to the integral can be estimated by expanding
\begin{align}U(r_++u)\approx \frac{4\pi}{\beta} u+\frac{d(d-1)}{\ell^2}u^2+\mathcal{O}(u)^3,\nonumber\\\end{align}
which yields
\begin{align}\left.\frac{2\pi \mathcal{A}}{\kappa_{d+1}}\right\vert_{\text{Near horizon}}\approx\frac{2\pi r_+^{d-1}}{\kappa_{d+1}}V_{d-2}\frac{\log(\beta \Phi/\ell)}{\sqrt{2}(d-2)\Phi},\label{areafu3}\end{align}
a logarithmically divergent result, \eq{ncon} of the main text.

\subsection{Entanglement entropy for spherical region}
We will first take the entangling surface to be a $S^{d-2}$-dimensional sphere of radius $L$ on the boundary $\mathbb{R}^d$ where the CFT lives. Using spherical coordinates on the $\mathbb{R}^{d-1}$ factor, with radial coordinate $\zeta$, we can parametrize the RT surface by the function $r(\zeta)$. The area functional is now
\begin{equation}\mathcal{A}=2\omega_{d-2}\int_0^L(r\,\zeta)^{d-2}\sqrt{\left(\frac{dr}{d\zeta}\right)^2+r^2}d\,\zeta.\end{equation}
This time, the equations of motion admit no conserved quantities. It is nevertheless still true that, as $R\rightarrow\infty$, the extremal surface sits at $r\approx r_+$ for a long time. Substituting $r\approx r+$, one gets again a volume law,
\begin{equation}\frac{2\pi \mathcal{A}}{\kappa_{d+1}} \approx  \frac{2\pi r_+^{d-1}}{\kappa_{d+1}} \frac{\omega_{d-2} L^{d-1}}{d-1},\end{equation}
which is equation \eq{lala}.

\section{Exponential decay of correlations at equal time}\label{app:lieb}
In this Appendix we prove the claim in Section \ref{sec:bhp}, that semi-local criticality at large Euclidean time together with vanishing of commutators at spacelike separation implies exponential decay of correlations at $t=t'$. We will essentially copy a similar derivation in \cite{2010arXiv1008.5137H}, which shows how a Lieb-Robinson bound implies exponential decay of correlations in gapped systems. 

Consider the two-point function of two operators  $\mathcal{O}$ and $\mathcal{O}'$. The two-point function at $t=t'=0$ is simply
\begin{equation}\langle\Psi\vert \mathcal{O}'(0,0)\mathcal{O}(\vec{x},0)\ket{\Psi}.\label{fin1}\end{equation}
Let us now introduce a new operator $\mathcal{O}^{+}(\vec{x})$, the positive-frequency part of $\mathcal{O}(\vec{x})$. That is, the matrix elements between eigenstates $\ket{\omega},\ket{\omega'}$ of energies $\omega,\omega'$ are 
\begin{equation} \langle \omega\vert  \mathcal{O}^{+}(\vec{x}) \ket{\omega'}\equiv  \langle \omega\vert \mathcal{O}(\vec{x}) \ket{\omega'} \theta(\omega-\omega').\end{equation}
The state $\ket{\Psi}$ which displays semi-local behavior is a ground state. There are no states of lower energy, so $\bra{\Psi} \mathcal{O}^{+}(\vec{x})=\bra{\Psi}\mathcal{O}(\vec{x})P_0$, where $P_0$ is the projector onto the groundstate subspace. Now, consider the momentum space matrix element 
\begin{equation}\bra{\Psi}\mathcal{O}(\vec{p},0) \ket{\Psi'}\label{1ptbm}\end{equation}
where $\ket{\Psi},\ket{\Psi'}$ are two elements of the ground state subspace. They are two different black brane microstates, or two thermofield double states with different phases. For single-trace operators of low near-horizon dimension, correlators  are insensitive to the particular black hole microstate under consideration, so we need only consider $\ket{\Psi}=\ket{\Psi'}$. We can argue that this holds more generally, but in a handwavy way. Since the black brane has infinite volume, one would naively expect different ground states to lie in different superselection sectors. Matrix elements of local operators such as \eq{1ptbm} should then vanish unless $\ket{\Psi}=\ket{\Psi'}$.  Equation \eq{1ptbm} scales with dimension $\Delta(\vec{p})$ under the near-horizon emergent scaling symmetry, so it must vanish (even when diagonal) as long as $\Delta(\vec{p})>0$. Then one gets $\bra{\Psi} \mathcal{O}^{+}(\vec{x})=0$ and $\mathcal{O}^{+}(\vec{x})\ket{\Psi}=\mathcal{O}(\vec{x})\ket{\Psi}$, so \eq{fin1} can be rewritten as
\begin{equation}C(\vec{x})\equiv\langle\Psi\vert  \mathcal{O}'(0,0)\mathcal{O}(\vec{x},0)\ket{\Psi}= \langle\Psi\vert  [\mathcal{O}'(0,0),\mathcal{O}^{+}(\vec{x},0)]\ket{\Psi}.\label{fin2}\end{equation}
We have turned the correlator into a commutator. One can further use the integral representation
\begin{equation}\tilde{\mathcal{O}}^{+}(\vec{x})=\frac{1}{2\pi}\lim_{\epsilon\rightarrow 0} \int_{-\infty}^\infty dt\, \frac{1}{it+\epsilon}\mathcal{O}(\vec{x},t),\end{equation}
which allows us to write the integral in terms of ordinary commutators, as
\begin{equation}C(\vec{x})=\frac{1}{2\pi}\lim_{\epsilon\rightarrow 0} \int_{\vert t\vert \geq \vert\vec{x}\vert} dt\, \frac{1}{it+\epsilon} \langle\Psi\vert  [\mathcal{O}'(0,0),\mathcal{O}(\vec{x},t)]\ket{\Psi}.\label{fin3}\end{equation}
We used the fact that in a Lorentzian QFT the commutator vanishes outside the lightcone to restrict the integration range\footnote{In \cite{2010arXiv1008.5137H}, a similar commutator in this region is instead exponentially suppressed, via the Lieb-Robinson bound.}. The Lorentzian commutator in \eq{fin3} can be rewritten in terms of a difference of Lorentzian two-point functions. As usual in QFT, the Euclidean 2-point function
\begin{equation}\langle \tilde{\mathcal{O}}(-\vec{p},0) \tilde{\mathcal{O}}(\vec{p},t_E)\rangle=\frac{\mathcal{N}(\vec{p})}{\vert t_E\vert^{2\Delta(\vec{p})}}\end{equation}
where $\tilde{\mathcal{O}}$ is the spatial Fourier transform of $\mathcal{O}$ and $\mathcal{N}(\vec{p})$ is a normalization factor, is related to the Lorentzian one via a Wick rotation, and in particular 
\begin{equation} \langle \tilde{\mathcal{O}}(-\vec{p},0) \tilde{\mathcal{O}}(\vec{p},t)\rangle= \frac{\mathcal{N}(\vec{p})e^{-i\pi \Delta(\vec{p}) \text{sign} (t)}}{\vert t\vert^{2\Delta(\vec{p})}}\end{equation}
which means
\begin{equation}\langle\Psi\vert  [\mathcal{O}'(0,0),\mathcal{O}(\vec{x},t)]\ket{\Psi}=2i \int d\vec{p}\,e^{i\vec{p}\cdot\vec{x}}\,  \frac{\mathcal{N}(\vec{p})\sin(\pi \Delta (\vec{p}))}{\vert t\vert^{2\Delta(\vec{p})}}.\label{fin4}\end{equation}
As explained in \cite{Iqbal:2011in,Hartnoll:2016apf,Gralla:2018xoz}, a holomorphic $\Delta(\vec{p})$ near zero, or more precisely on a strip near the real axis, leads to exponential decay in \eq{fin4}. Technically, this is a consequence of the Paley-Wiener theorem (see e.g. \cite{reed1975ii}, Theorem IX.13). The normalization factor $\mathcal{N}(\vec{p})$ is also holomorphic on a strip near zero, since we are assuming semi-local behavior for $t\gg 1/\mu$. All in all, for large enough $t$,
\begin{equation}\vert \langle\Psi\vert  [\mathcal{O}'(0,0),\mathcal{O}(\vec{x},t)]\ket{\Psi}\vert \leq \frac{e^{-\xi \vert\vec{x}\vert}}{\vert t\vert^{2\Delta_{\text{min}}}}.\end{equation}
Here, $\xi$ is some correlation length (given by the width of the strip in the complex $p^2$ plane where  $\Delta(\vec{p})$ is analytic), and $\Delta_{\text{min}}>0$ is the minimum value of $\Delta(\vec{p})$ for real argument. Plugging this back into \eq{fin3}, and using $\Delta_{\text{min}}>0$ so that the time integral is convergent, one obtains $\vert C(\vec{x})\vert\leq e^{-\xi \vert\vec{x}\vert}$, as advertised.

If one introduces a lattice regulator, as we do in the main text, then Lorentz invariance is broken explicitly and the commutator no longer vanishes at spacelike separation. However, a Lieb-Robinson bound \cite{Lieb:1972wy,2010arXiv1008.5137H} guarantees that, as long as interactions are local, there is an effective lightcone outside of which the commutator vanishes exponentially fast. Therefore, the result still holds. 

This argument establishes exponential decay of correlations at equal times for one copy of the CFT, but the argument can be extended to two-sided correlators in the thermofield double formalism. A generic local operator in this case is of the form $\mathcal{O}_L(\vec{x},t) \mathcal{O}_R(\vec{x},t)$, where the left and right insertions are at the same point $\vec{x}$ in both CFT's. The two-point function  in the ground state thermofield double \eq{tfdt0} is
\begin{align} &\langle TFD\vert \mathcal{O}_L(\vec{p},t) \mathcal{O}_R(\vec{p},t) \mathcal{O}_L(0,0) \mathcal{O}_R(0,t)\ket{TFD}\nonumber\\&=\sum_{\text{Groundstates}} \langle a'\vert_L \mathcal{O}_L(\vec{p},t) \mathcal{O}_L(0,0) \ket{a}_L \langle \bar{a'}\vert_R \mathcal{O}_R(\vec{p},t) \mathcal{O}_R(0,0)\ket{\bar{a}}_R,\label{2sided}\end{align}
so the two-point function is controlled by the off-diagonal matrix elements 
\begin{equation} \langle a'\vert \mathcal{O}_L(\vec{p},t) \mathcal{O}_L(0,0) \ket{a}\propto \frac{\mathcal{O}_{aa'}}{\vert t\vert^{2\Delta_{\mathcal{O}}(\vec{p})}}\label{2sided2}\end{equation}
between different black brane microstates. Here, $\mathcal{O}_{aa'}$ is some unknown matrix element; the time and momentum dependence is fixed by the near-horizon scaling symmetry and the fact that any ground state is invariant under it. For single-trace operators, this is again the statement that one-sided correlators only depend on the exterior geometry. Fourier-transforming the two-sided correlator \eq{2sided} back to real space, we see that it inherits the exponential decay of correlations from the one-sided operators.

\bibliographystyle{jhep}
\bibliography{EEwgc}

\end{document}